\makeatletter
\declare@file@substitution{revtex4-1.cls}{revtex4-2.cls}

\makeatother
\documentclass[%
 reprint,
superscriptaddress,
 amsmath,amssymb,
 aps, physrev,
]{revtex4-2}

\usepackage{graphicx}        
\usepackage{xcolor}         
\usepackage{amsmath,amssymb}
\usepackage{physics}
\usepackage{hyperref}

\newcommand{\dif}{\mathrm{d}}

\begin{document}

\title{Enhanced All-Distance Equi-Zenith Angle Method for Cosmic-Ray Anisotropy Measurement}

\author{Dong-Xu Sun}
\affiliation{Key Laboratory of Dark Matter and Space Astronomy, Purple Mountain Observatory, Chinese Academy of Sciences, Nanjing 210023, China}
\affiliation{School of Astronomy and Space Science, University of Science and Technology of China, Hefei 230026, China}

\author{Dan Li}
\affiliation{Key Laboratory of Particle Astrophysics, Institute of High Energy Physics, Chinese Academy of Sciences, Beijing 100049, China}
\affiliation{University of Chinese Academy of Sciences, Beijing 100049, China}

\author{Wei Liu}
\email{liuwei@ihep.ac.cn}
\affiliation{Key Laboratory of Particle Astrophysics, Institute of High Energy Physics, Chinese Academy of Sciences, Beijing 100049, China}
\affiliation{Tianfu Cosmic Ray Research Center, 610000 Chengdu, Sichuan, China}

\author{Qiang Yuan}
\affiliation{Key Laboratory of Dark Matter and Space Astronomy, Purple Mountain Observatory, Chinese Academy of Sciences, Nanjing 210023, China}
\affiliation{School of Astronomy and Space Science, University of Science and Technology of China, Hefei 230026, China}

\author{Yi-Qing Guo}
\affiliation{Key Laboratory of Particle Astrophysics, Institute of High Energy Physics, Chinese Academy of Sciences, Beijing 100049, China}
\affiliation{University of Chinese Academy of Sciences, Beijing 100049, China}
\affiliation{Tianfu Cosmic Ray Research Center, 610000 Chengdu, Sichuan, China}

\author{Hong-bo Hu}
\affiliation{Key Laboratory of Particle Astrophysics, Institute of High Energy Physics, Chinese Academy of Sciences, Beijing 100049, China}
\affiliation{University of Chinese Academy of Sciences, Beijing 100049, China}
\affiliation{Tianfu Cosmic Ray Research Center, 610000 Chengdu, Sichuan, China}

\begin{abstract}
Long-term observations indicate that the relative intensity of cosmic-ray anisotropy remains below $0.1\%$ for energies less than $\sim 1$ PeV. Measuring such faint signals poses a significant challenge in data analysis, requiring careful removal of instrumental and atmospheric artifacts. The all-distance equi-zenith angle method is widely employed to extract cosmic-ray anisotropies, as it effectively suppresses the instantaneous variations arising from the instrument and atmosphere. \textcolor{black}{However, instability in the detector efficiency makes precise measurements of anisotropy challenging with this method.} In this work, we present an enhanced all-distance equi-zenith angle method for cosmic-ray anisotropy measurement. Unlike previous implementations, our improved approach enables the simultaneous measurement of anisotropies over multiple time frames and allows the detection efficiency to be determined directly from the data. This feature makes the method especially suitable for applications where the detector array does not operate with long-term stability\textcolor{black}{, and thus allows for the measurement of anisotropy with high-precision}. Moreover, our enhanced method is also feasible when the data do not span complete tropical years.
\end{abstract}

\maketitle

\section{Introduction}

With the exception of the ultra-high-energy regime, most cosmic rays (CR) undergo frequent scattering off turbulent magnetic fields during propagation, leading to a highly isotropic distribution of their arrival directions. However, numerous observations have identified weak but statistically significant anisotropies on the order of about $0.1 \%$. Although space-based detectors offer the advantage of full-sky coverage, their limited detection area makes it challenging to measure such faint signals. Currently, CR anisotropies are mainly measured by the ground-based experiments. To date, CR anisotropies have been detected across a broad energy range from $\sim$ GeV to beyond EeV, with relative intensities on the order of approximately $10^{-4}$ to $10^{-2}$ \citep{2006Sci...314..439A, 2007PhRvD..75f2003G, 2008PhRvL.101v1101A, 2009ApJ...698.2121A, 2010ApJ...718L.194A, 2010ApJ...711..119A, 2011ApJ...740...16A, 2012ApJ...746...33A, 2013ApJ...765...55A, 2013PhRvD..88h2001B, 2015ApJ...809...90B, 2014ApJ...796..108A, 2016ApJ...826..220A, 2017ApJ...836..153A, 2017Sci...357.1266P, 2019ApJ...871...96A, 2022ApJ...938...30M, 2025arXiv251218401L, 2026arXiv260102801C}. The available anisotropies contain sidereal and solar diurnal modulations, as well as transient and steady signals. Moreover, the anisotropies exhibit a strong energy dependence and manifest across a wide range of angular scales.

Compared with the energy spectrum and composition, the CR anisotropy provides additional insights into their source distribution and propagation conditions. Below approximately $1$ TeV, the anisotropy is strongly modulated by the intense solar activities, such as solar flares and interplanetary coronal mass ejections \citep{2022ApJ...938...30M, 2026arXiv260102801C}. Between about $1$ TeV and $1$ PeV, the dipole anisotropy points to the presence of a local CR accelerator \citep{2016PhRvL.117o1103A, 2019JCAP...10..010L, 2019JCAP...12..007Q, 2021PhRvD.104j3013F, 2022MNRAS.511.6218Z}, and show a close connection with the spectral hardening observed above $200$ GeV. Moreover, the dipole phase suggests that CRs experience anisotropic diffusion in the local region before entering the heliosphere \citep{2014Sci...343..988S, 2016PhRvL.117o1103A, 2020ApJ...892....6L}. At energies above EeV scale, a dipole modulation oriented away from the Galactic center has been detected, supporting an extragalactic origin for CRs at these energies. Medium- and small-scale anisotropies have been observed at levels around $10^{-4}$ at TeV energies. The origin of these structures is linked to CR propagation in the local interstellar medium or the influence of the heliosphere. For a comprehensive review of observations and theoretical explanations at these scales, one can refer to \citep{2017PrPNP..94..184A}.

Despite growing interest in CR anisotropy, its measurement remains a highly challenging task. The signal is exceedingly weak, requiring detectors with large collecting areas and sustained stable operation to achieve adequate statistics. At energies above $\sim 1$ TeV, CRs are detected indirectly via extensive air showers (EAS) using ground-based arrays. They measure anisotropies along the right ascension by leveraging the Earth's rotation. However, the spurious modulations in counting rates inevitably arise from both instrumental (e.g., data-taking interruptions or acceptance changes) and meteorological effects (e.g., temperature and pressure influences on EAS development), which hamper the extraction of the faint anisotropy signal. Therefore, the precise correction for subtle variations in detection efficiency is required. One may appeal to the accurate simulation of the detector response. But that is difficult to achieve in practice.

Given these experimental constraints, the analysis of CR anisotropy necessitates methods that can estimate or correct for detection efficiency directly from the data itself, minimizing reliance on detector simulation. Over the past decades, the anisotropy measurements have driven the development of various such techniques. Prior to the 2000s, anisotropy studies were primarily limited to one-dimensional harmonic analysis in the right ascension, employing methods such as the Rayleigh formalism \citep{1975PhRvL..34.1530L} or the East-West approach \citep{2011ApJ...738...67B}. Since then, advancements in detection technology, yielding larger statistics and improved angular resolution, have enabled two-dimensional morphological studies of CR arrival directions. Notable techniques developed for this purpose include the all-distance equi-zenith angle method \citep{2005ApJ...633.1005A}, the forward-backward method \citep{2009ApJ...698.2121A}, the direct integration method \citep{2003ApJ...595..803A}, the time-scrambling technique \citep{1993NIMPA.328..570A}, the maximum-likelihood method \citep{2016ApJ...823...10A}, and so on.

The all-distance equi-zenith angle method has been successfully used to reconstruct CR anisotropies in experiments such as AS$\gamma$ \citep{2005ApJ...633.1005A, 2006Sci...314..439A, 2017ApJ...836..153A}, ARGO \citep{2015ApJ...809...90B, 2018ApJ...861...93B}, and LHAASO \citep{2024icrc.confE.186L, 2024icrc.confE.478G, 2025icrc.confE.320L, 2025icrc.confE.266Z, 2025arXiv251218401L}. In this approach, the background for a given ``on-source bin" is estimated using data from the sideband region of the same zenith-angle belt observed at the same time. This effectively mitigates the potential detection effects arising from instantaneous environmental variations that are difficult to monitor in real time and could otherwise introduce systematic errors into the measurement.
However, the original method performs optimally when the detection efficiency is nearly independent of azimuth. Otherwise, a dedicated correction for azimuth-dependent response has to be performed prior to the analysis. And the anisotropies of different time frames (e.g., sidereal, solar) are derived separately. In this study, we present an enhanced version of the all-distance equi-zenith angle method. Compared to the original approach, our improved method enables the simultaneous measurement of anisotropies across multiple time frames and allows the detection efficiency to be determined from the data itself. This advancement makes the method particularly well-suited for applications when detector arrays are not in a fully stable operational state. Moreover, the new method does not require data to span complete tropical years to mitigate the interference from other time frames.

This paper is organized as follows. Section 2 introduces the baseline all-distance equi-zenith angle method and details the enhancements we have developed. Section 3 is devoted to verifying the performance of our improved method through Monte Carlo simulations. The summary and concluding remarks are provided in Section 4.

\section{Enhanced All-Distance Equi-Zenith Angle Method}

\subsection{Introduction to four time frames}

\textcolor{black}{
At present, anisotropies are measured in four time frames: sidereal, solar, anti-sidereal, and extended-sidereal. The physical origins of anisotropies have been observed in the sidereal and solar time frames, respectively. The primary anisotropy arises from the inhomogeneous distribution of distant cosmic-ray (CR) sources and their subsequent diffusive propagation in the Galaxy. This anisotropy is observed in the sidereal time frame, which uses a fixed star as a reference point, yielding a sidereal day of $T_{\rm sid} = 23^{\rm h} 56^{\rm m} 4^{\rm s}$. The sidereal anisotropy varies with a period of one sidereal day, corresponding to an angular frequency of $\omega_{\rm sid} = 1/T_{\rm sid}$.
}

\textcolor{black}{
A second type of anisotropy arises as a kinematic effect due to the Earth's orbital motion around the Sun. This effect predicts a dipole anisotropy in the local solar time frame and is commonly referred to as the solar Compton–Getting (CG) effect \citep{1935PhRv...47..817C, 1986Natur.322..434C}. Because its amplitude and phase are independent of energy, the solar CG effect is usually regarded as a ``standard candle" in CR anisotropy measurements. Solar time is measured by Earth's rotation relative to the Sun, with one solar day lasting $T_{\rm sol} = 24^{\rm h}$. Likewise, the solar anisotropy varies with a period of one solar day, corresponding to an angular frequency of $\omega_{\rm sol} = 1/T_{\rm sol}$.
}

\textcolor{black}{
At any given time, the observed CR arrival distribution contains both sidereal and solar anisotropies. When the observation time $T$ exceeds $2\pi / \Delta \omega$, where $\Delta \omega$ is the difference in angular frequency, the signals at different angular frequencies can be decoupled from each other \citep{2019APh...104...13D}. Since a sidereal day is about $4$ minutes shorter than a solar day, when the observation period exceeds one tropical year, the sidereal and solar anisotropies can be naturally separated. In this case, when data from full years are binned in local sidereal time, the solar anisotropy is expected to average out, and vice versa.
}

\textcolor{black}{
However, when the CR signal is modulated by a long-term variations, such as temperature, pressure, etc., the anisotropy at one frequency can induce signals at adjacent frequencies \citep{1954PPSA...67..996F, 1954Natur.173..445F, 1957PPSB...70..840F}. Specifically, when the solar anisotropy is modulated by a seasonal effect, a signal would be induced at the sidereal and anti-sidereal angular frequencies; likewise, when the sidereal anisotropy is modulated by a seasonal effect, a signal would be induced at the solar and extended-sidereal frequencies. The anti-sidereal day is longer than the solar day by $4$ minutes, while the extended-sidereal day is $4$ minutes shorter than the sidereal day. Since there are no physical effects associated with either time frame, no signal is expected to be detected in either frame. Therefore, anisotropy measurements in the anti-sidereal and extended-sidereal time frames can verify whether the sidereal(solar) anisotropy is contaminated by the solar(sidereal) anisotropy and can used to estimate the corresponding systematics in sidereal(solar) time frame.
}

\subsection{Description of Conventional All-Distance Equi-Zenith Angle Method}

\textcolor{black}{
The anisotropies in the four time frames are analyzed separately using the original all-distance equi-zenith angle method \citep{2005ApJ...633.1005A, 2012APh....39..144L}, which is outlined as follows. Here, we take the analysis of sidereal anisotropy as an example.
}

\begin{enumerate}

\item \textcolor{black}{To measure the anisotropy in sidereal time, the observation time is divided according to the local sidereal time. At a given local sidereal time $t$, the field of view is divided into spatial bins of equal solid angle, based on local coordinates. The CR events recorded over full years are then assigned to bins $N(t, \theta, \phi)$ according to both the local sidereal time $t$ and their arrival directions (zenith angle $\theta$ and azimuth $\phi$).
}

\item \textcolor{black}{Given the stable non-uniform azimuthal efficiency of the detector array, an azimuthal correction is applied to all data bins. For each zenith angle belt, the azimuthal distribution accumulated over the full tropical year is normalized so that its mean value is $1$. By inverting this normalized distribution and using it as an event weight, the total number of events is preserved while removing any non-uniformity along the azimuthal angle.
}

\item \textcolor{black}{In the equal-zenith-angle method, when a given bin is taken as an ``on-source" bin with an event count of $N_{\rm on}(t, \theta, \phi)$, the background is estimated using events from the other bins within the same zenith belt at the same local sidereal time $t$,} That is,  
\begin{align}
\dfrac{1}{n_\theta -1} \sum\limits_{\phi^\prime \neq \phi} N_{\rm off}(t, \theta, \phi^\prime) ~,
\end{align}
where $N_{\rm off}(t, \theta, \phi^\prime)$ is the event count of a single ``off-source" bin, $n_\theta$ is the number of bins in the zenith angle belt $\theta$, and the sum runs over all bins in the same zenith belt except the ``on-source" bin itself. To extract the anisotropy, a least-squares method is applied, and a $\chi^2$ statistic can be constructed as follows:
\begin{equation}
\chi^2 = \sum_{t,\theta,\phi} \dfrac{\left[ \dfrac{N_{\text{on}}(t, \theta, \phi)}{I(\delta, \alpha)} - \dfrac{1}{n_{\theta}-1}\sum\limits_{\phi^\prime \neq \phi}  \dfrac{N_{\text{off}}(t, \theta, \phi')}{I(\delta^\prime, \alpha^\prime)} \right]^2 }{\sigma^2(t, \theta, \phi)} ~.
\end{equation}
\textcolor{black}{$I(\delta,\alpha) = 1 +\delta I(\delta,\alpha)$ is the relative intensity in the equatorial coordinates (declination $\delta$ and right ascension $\alpha$), with $\delta I(\delta,\alpha)$ denoting the small deviation from isotropy. It is connected, via a time-dependent coordinate transformation, to each grid cell $N(t,\theta,\phi)$ in the local coordinate system.} The variance $\sigma^2$ incorporates uncertainties from both ``on-source" and ``off-source" bins, i.e.
\begin{equation}
\sigma^2(t, \theta, \phi) = \dfrac{N_{\text{on}}(t, \theta, \phi)}{I^2(\delta, \alpha)} + \frac{1}{n_{\theta}-1}\sum_{\phi^\prime \neq \phi} \dfrac{N_{\text{off}}(t, \theta, \phi^\prime)}{I^2(\delta^\prime, \alpha^\prime)} ~,
\end{equation}
which can be derived from the error propagation. The summation is performed over all grids in the field of view for one local sidereal day. Every bin serves alternately as ``on-source" and ``off-source". 
\item \textcolor{black}{
When the $\chi^2$ statistic reaches its minimum, i.e., $\dfrac{\partial \chi^2}{\partial I(\delta, \alpha)} = 0$ for every $I(\delta, \alpha)$, the two-dimensional distribution of the relative intensity of the anisotropy can be determined. To solve such complicated equations, an iterative method is introduced into the $\chi^2$ statistic, i.e.}
\begin{align}
\color{black} \chi^2 = \sum_{t,\theta,\phi} \dfrac{\left[ \dfrac{N_{\text{on}}(t, \theta, \phi)}{I^{k+1}(\delta, \alpha)} - \dfrac{1}{n_{\theta}-1}\sum\limits_{\phi^\prime \neq \phi}  \dfrac{N_{\text{off}}(t, \theta, \phi')}{I^k(\delta^\prime, \alpha^\prime)} \right]^2 }{[\sigma^k(t, \theta, \phi) ]^2} ~,
\end{align}
\textcolor{black}{in which $I^{k+1}(\delta,\alpha)$ is the result at the $k+1$-th iteration, while $I^{k}(\delta,\alpha)$ and $\sigma^k(t,\theta,\phi)$ are those from the previous iteration. The estimator at the $k+1$-th iteration, $I^{k+1}(\delta, \alpha)$, can be expressed as}
\begin{align}
\color{black} I^{k+1}(\delta,\alpha) = \dfrac{\sum\limits_{t, \theta, \phi} N_{\text{on}}(t,\theta,\phi)}{\sum\limits_{t, \theta, \phi} \dfrac{1}{n_\theta-1} \sum\limits_{\phi^\prime \neq \phi} \dfrac{N_{\text{off}}(t,\theta,\phi^\prime)}{I^k(\delta^\prime, \alpha^\prime)} } ~.
\label{eq:iter_I}
\end{align}
\textcolor{black}{The initial values, $I^0(\delta,\alpha)$, are set to unity. Throughout the iteration process, the global normalization constraint $\dfrac{1}{4\pi} \displaystyle \int I(\delta, \alpha) \, \mathrm{d} \Omega = 1$ or $\displaystyle \int \delta I(\delta, \alpha) \, \mathrm{d} \Omega = 0$ is enforced to prevent divergence. After sufficient iterations, stable solutions for the sidereal anisotropy $I(\delta, \alpha)$ and its corresponding uncertainty $\sigma(\delta, \alpha)$ are obtained. Since conventional ground-based observatories are insensitive to relative intensity variations across declination bands, the reconstructed anisotropy is renormalized for each declination band after iteration, i.e. $\dfrac{1}{2\pi} \displaystyle\int I(\delta, \alpha) \, \mathrm{d} \alpha = 1$.}
\end{enumerate}
\textcolor{black}{The same analysis procedure can be applied to determine the anisotropies in the other three time frames.}

\subsection{Improvement of Analysis Method}

\textcolor{black}{However, the above method requires observational data spanning complete years to average out signals from other frequencies when filling CR events into spatial grids in local time during the first step. More importantly, if the azimuthal efficiency becomes unstable due to instrumental instability, a more careful correction of variable detection efficiencies must be performed.}

\textcolor{black}{
The detection efficiency exhibits both zenith and azimuthal dependence. As an extensive air shower develops in the atmosphere, its lateral and longitudinal evolution is influenced by the atmospheric state \citep{2009APh....32...89P}. The number of detected events decreases sharply with increasing zenith angle, which is approximately expressed as
\begin{align}
\dfrac{\dif R}{\dif \cos \theta} \propto \cos^n \theta \exp \left[ -\dfrac{KT}{\cos \theta} \right]
\end{align}
where $\cos^n$ results from the detector response \citep{1993NIMPA.328..570A}. In the all-distance equi-zenith angle method, the background is estimated using events from the sideband regions within the same zenith belt and at the same observation time. This approach cancels out the zenith-dependent detection efficiency variations, as can be seen from Equ. (\ref{eq:iter_I}). Therefore, only the azimuthal efficiency remains a concern.
}

\textcolor{black}{
The azimuthal dependence of the detection efficiency may arise from trigger biases due to the specific geometric layout of the array. Operations such as detector cleaning or electronics restarting may cause the azimuthal dependence of the detection efficiency to change sharply. These variations in the azimuthal distribution are difficult to correct using the azimuthal correction method and may interfere with precise measurements of anisotropy.
}

\textcolor{black}{
Additionally, the geomagnetic field can also induce azimuthal modulations. On one hand, the geomagnetic field deflects low-rigidity CRs, giving rise to the well-known east–west asymmetry. On the other hand, it affects the trajectories of secondary charged particles in extensive air showers, stretching their lateral distribution. This effect depends on the angle between the shower axis and the geomagnetic field, thereby introducing non-uniformity in the azimuthal distribution \citep{2014PhRvD..89e2005B}. The azimuthal distribution caused by the geomagnetic effect can be considered stable. However, this effect cannot be disentangled from the detector efficiency by the current approach and is collectively treated as part of the detector efficiency.
}

\textcolor{black}{
In the enhanced method, CR events are binned into consecutive Modified Julian Days (MJDs) rather than local time. In this case, the anisotropies at sidereal frequency do not average out. Since the anisotropy at each frequency can be considered an independent process, the total relative intensity $I$ can be expressed as the product of anisotropies across multiple distinct time frames. Furthermore, if the azimuthal efficiency is not corrected, it can also be included as a factor in the product. The total relative intensity $I$ can be written as
\begin{align}
I &= I_{\rm sid}(\delta_{\rm sid},\alpha_{\rm sid}) 
     \times I_{\rm sol}(\delta_{\rm sol},\alpha_{\rm sol}) \notag\\
  &\quad \times I_{\rm asid}(\delta_{\rm asid},\alpha_{\rm asid}) 
     \times I_{\rm esid}(\delta_{\rm esid},\alpha_{\rm esid}) 
     \times \varepsilon(t,\theta,\phi) ~.
\label{eq:I_comb}
\end{align}
To estimate the systematics, the anti-sidereal ($I_{\rm asid}$) and extended-sidereal ($I_{\rm esid}$) components are also included. The detection efficiency $\varepsilon$ is expressed in local coordinates $(\theta,\phi)$ and may vary with time. The coordinates $\delta$ and $\alpha$ for each frequency component can be connected to each grid cell $N(t,\theta,\phi)$ in the local coordinate system via separate time-dependent coordinate transformations. The estimators of sidereal and solar anisotropies at the $k+1$-th iteration, can be written as
\begin{align}
I^{k+1}_{\rm sid} = \dfrac{\sum\limits_{t,\theta,\phi} N_{\text{on}} }{\sum\limits_{t,\theta,\phi} \dfrac{ \dfrac{1}{n_\theta-1} \sum\limits_{\phi^\prime \neq \phi} N_{\text{off}} \beta^k_{\rm sid}\beta^k_{\rm sol}\beta^k_{\rm asid}\beta^k_{\rm esid} \eta^k }{\beta^k_{\rm sol}\beta^k_{\rm asid}\beta^k_{\rm esid}\eta^k} } ~,
\label{eq:Iiter_comb_sid}
\end{align}
\begin{align}
I^{k+1}_{\rm sol} = \dfrac{\sum\limits_{t,\theta,\phi} N_{\text{on}} }{\sum\limits_{t,\theta,\phi} \dfrac{ \dfrac{1}{n_\theta-1} \sum\limits_{\phi^\prime \neq \phi} N_{\text{off}} \beta^k_{\rm sid}\beta^k_{\rm sol}\beta^k_{\rm asid}\beta^k_{\rm esid} \eta^k }{\beta^k_{\rm sid}\beta^k_{\rm asid}\beta^k_{\rm esid}\eta^k} } ~,
\label{eq:Iiter_comb_sol}
\end{align}
in which $\beta^k = 1/I^k$ and $\eta^k = 1/\varepsilon^k$. The remaining anisotropies and the detection efficiency can be written likewise. In each iteration, the four anisotropy components and the detection efficiency are derived sequentially, with only one anisotropy map being iterated at a time while the other three are held fixed. Specifically, first, $I^{k+1}_{\rm sid}$ is evaluated by inserting $(I^{k}_{\rm sid}, I^{k}_{\rm sol}, I^{k}_{\rm asid}, I^{k}_{\rm esid}, \varepsilon^{k})$ into the right-hand side of Eq. (\ref{eq:Iiter_comb_sid}); second, $I^{k+1}_{\rm sol}$ is evaluated by inserting $(I^{k+1}_{\rm sid}, I^{k}_{\rm sol}, I^{k}_{\rm asid}, I^{k}_{\rm esid}, \varepsilon^{k})$ into the right-hand side of Eq. (\ref{eq:Iiter_comb_sol}); third, $I^{k+1}_{\rm asid}$ is evaluated by inserting $(I^{k+1}_{\rm sid}, I^{k+1}_{\rm sol}, I^{k}_{\rm asid}, I^{k}_{\rm esid}, \varepsilon^{k})$ into the analogous equation, and so on. In this way, all four anisotropy components and the detection efficiency can be solved for simultaneously. During iteration, the global normalization for each anisotropy component must be enforced. The initial values of the detection efficiency $\varepsilon^0(\theta, \phi)$ are set to $1$.
}

\section{Monte Carlo Simulation and Performance}
To validate the enhanced all‑distance equi‑zenith angle method, Monte Carlo simulations are carried out. In the simulations, CR Events are binned in units of MJD, with an average daily event count set to $10^9$. Every day is divided into discrete time bins with a duration of $16$ minutes. Within each time bin, the sky is partitioned into equal‑solid‑angle grids of $2^\circ \times 2^\circ$ in the horizontal coordinate system, covering zenith angles up to $40^{\circ}$, which yields $1218$ spatial grids within the field of view. The number of events in each spatial grid are sampled from a Possion distribution based on the expected event count. In the equatorial coordinate system, declination and right ascension are binned separately with a resolution of $4^\circ$ per grid. The simulation site adopts the geographic location of the LHAASO experiment, with latitude $29.4^{\circ}$ N and longitude $100.1^{\circ}$ E.

In the simulations, both sidereal and solar anisotropies are incorporated. For the sidereal component, a dipole anisotropy with an amplitude of $1 \times 10^{-3}$ is assumed, which is comparable with current measurements at TeV energies. Its phase is directed toward equatorial coordinates $\delta = 30^{\circ}$, $\alpha = 200^{\circ}$. The amplitude of the solar Compton–Getting effect is given by $(2+\gamma)\beta$, where $\gamma$ is the spectral index of the CR energy spectrum and $\beta$ is the CR velocity in units of the speed of light. With a spectral index of $\gamma = 2.7$ and Earth's mean orbital velocity of about $30$ km/s, the expected amplitude of the solar Compton–Getting effect is $4.477 \times 10^{-4}$. The orientation of this dipole aligns with Earth's velocity vector, resulting in a maximum intensity near local dawn.

\subsection{Simultaneous fitting with stable detection efficiency}

\begin{figure*}[!htb]
  \centering
  \includegraphics[width=1.0\textwidth]{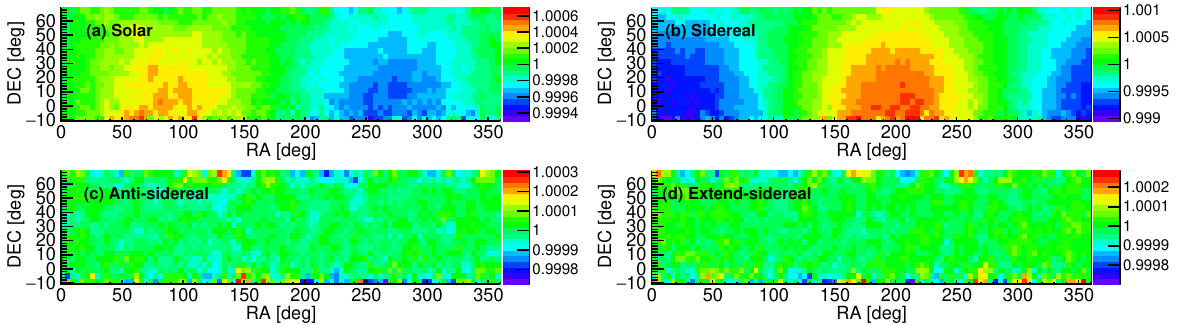}
  \caption{Anisotropy sky maps of the four time frames derived by the enhanced method. Panels (a)--(d) correspond to the solar, sidereal, anti-sidereal, and extended-sidereal time, respectively. The map is smoothed with a $5^\circ$ radius top-hat function.}
  \label{fig:4period_2d}
\end{figure*}

First, the enhanced method is applied to the scenario where the observation period is a full tropical year and a stable detection efficiency, assuming the input detection efficiency takes the form of
\begin{equation}
\varepsilon(\theta, \phi) = \cos^{8.5} \theta \sin \theta \exp \left[ -\dfrac{7.6}{\cos \theta} \right] \times [1+0.1 \sin(\phi)] ~.
\label{eq:effi_stable}
\end{equation}
The anisotropy sky maps for four distinct time frames—sidereal, solar, anti-sidereal, and extended-sidereal along with the detection efficiency, are derived simultaneously. Figure (\ref{fig:4period_2d}) presents the resulting anisotropy skymaps: solar (upper left), sidereal (upper right), anti-sidereal (lower left), and extended-sidereal (lower right). Each skymap has been smoothed with a top-hat function of $5^\circ$ radius. As anti-sidereal and extended-sidereal time frames are unphysical coordinate systems, they are not expected to contain anisotropies beyond random statistical fluctuations. Consistent with this expectation, the corresponding maps in the figure indeed exhibit only such fluctuations.

\begin{figure*}[!t]
  \centering
  \includegraphics[width=0.9\textwidth]{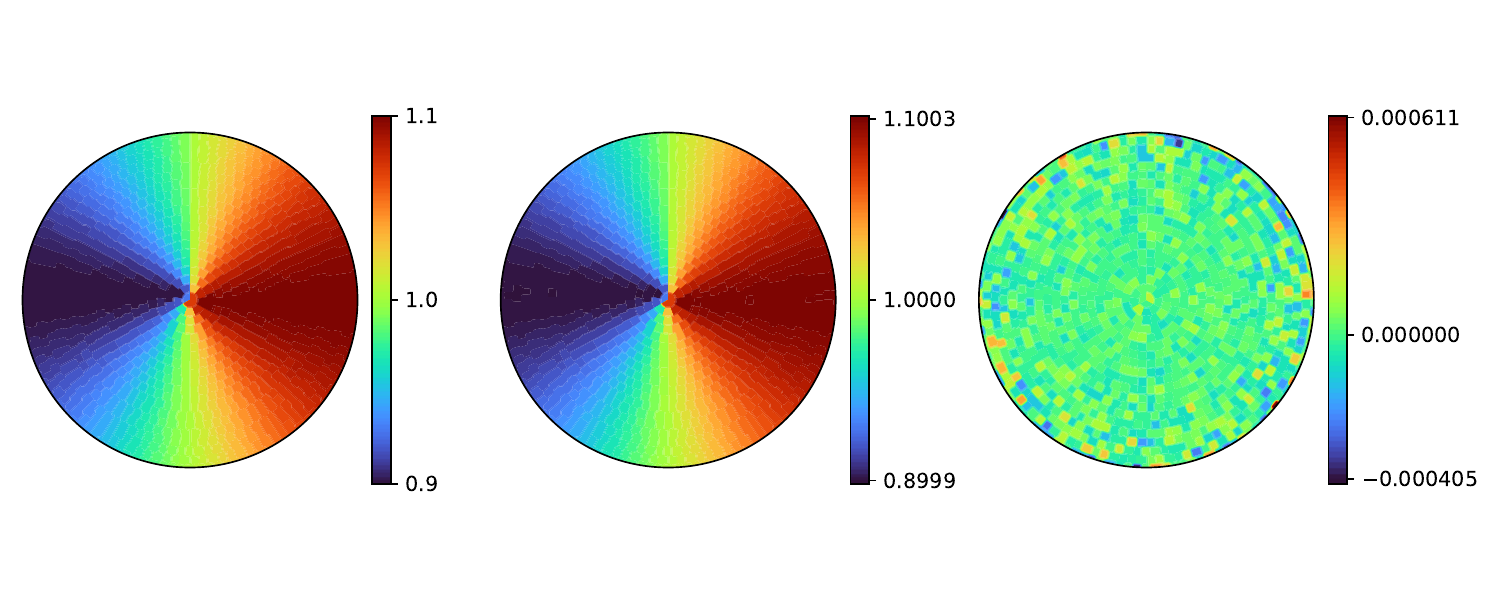}
  \caption{
Comparison of the expected and iterated efficiencies. The left panel shows the distribution of expected efficiency, the middle panel shows the iterated efficiency, and the right panel presents the residual map between them. Each skymap is centered at the zenith and extends to a zenith angle of $40^\circ$. The vertical direction points north. Following the clockwise direction, the sequence of azimuthal angles is east, south, and west.
  }
  \label{fig:eff_stable_dis_2d}
\end{figure*}

The sidereal and solar maps clearly exhibit dipole patterns, both centered at a declination of $0^\circ$. This is a consequence of the declination normalization applied, since the current ground-based measurement is insensitive to variations in anisotropy across different declination bands. In other words, this processing step effectively filters out the $m = 0$ multipole moments in the spherical harmonic expansion of the anisotropy. As a result, the dipole anisotropy is projected onto the celestial equator. The right ascension of the excess is approximately $90^\circ$ in the solar map and $200^\circ$ in the sidereal map, both of which are close to the expected values. \textcolor{black}{Figure (\ref{fig:eff_stable_dis_2d}) compares the iterated and input efficiency skymaps in the local coordinate. The left panel shows the expected efficiency, the middle panel shows the iterated efficiency, and the right panel presents the residual map between them. Because the zenith dependence of the detection efficiency is unknown, each azimuthal belt is normalized independently. The residual map exhibits fluctuations but no systematic anomalies, suggesting a good agreement between the iterated and expected efficiencies.}

\begin{figure*}[!htb]
  \centering

  \begin{minipage}[b]{0.49\textwidth}
    \centering
    \includegraphics[width=\textwidth]{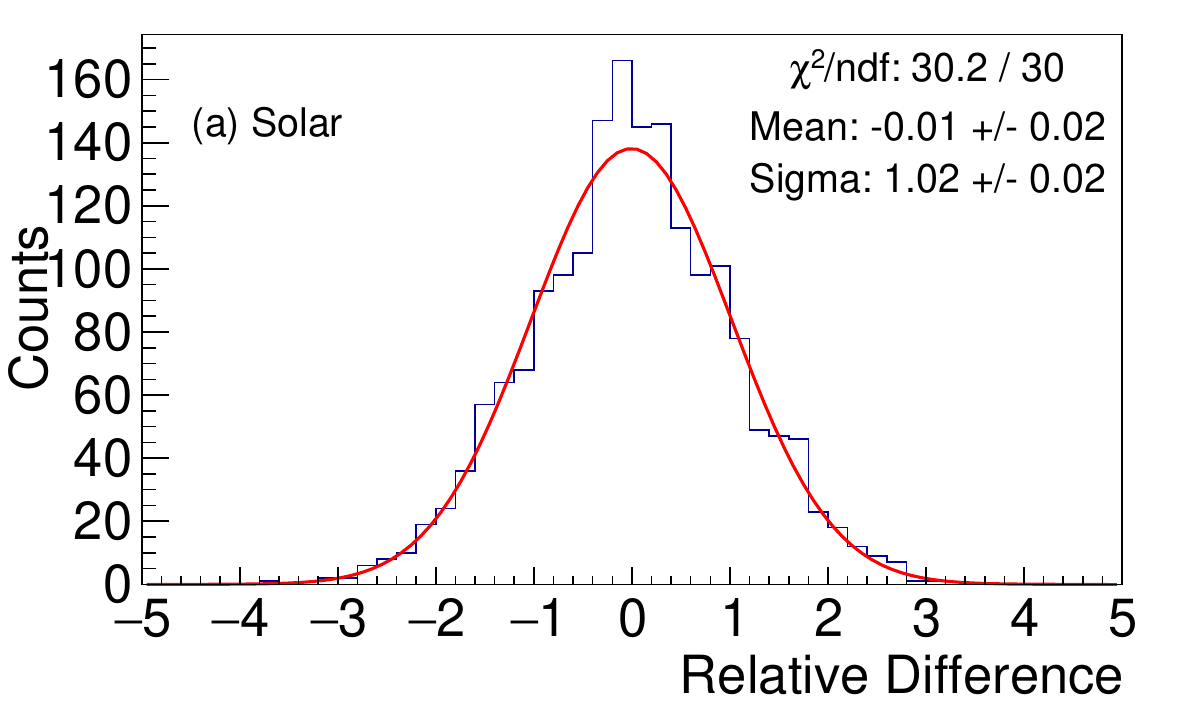}
  \end{minipage}%
  \hfill
  \begin{minipage}[b]{0.49\textwidth}
    \centering
    \includegraphics[width=\textwidth]{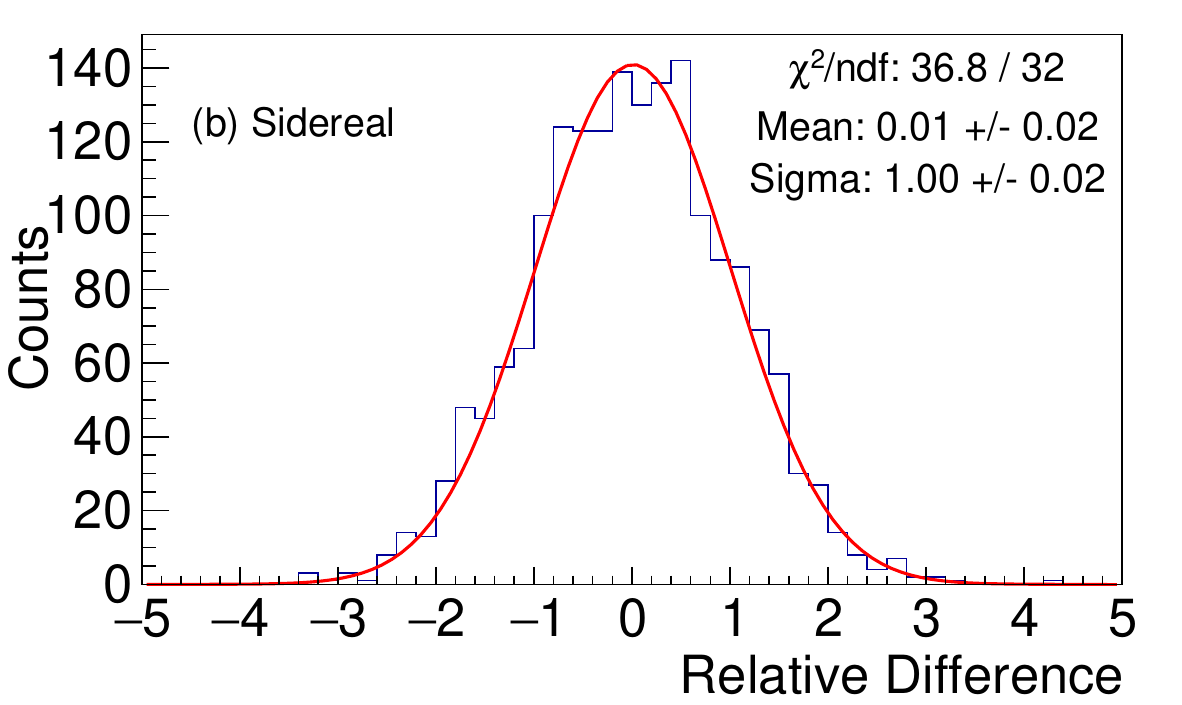}
  \end{minipage}

  \vspace{0.5em}
  \begin{minipage}[b]{0.49\textwidth}
    \centering
    \includegraphics[width=\textwidth]{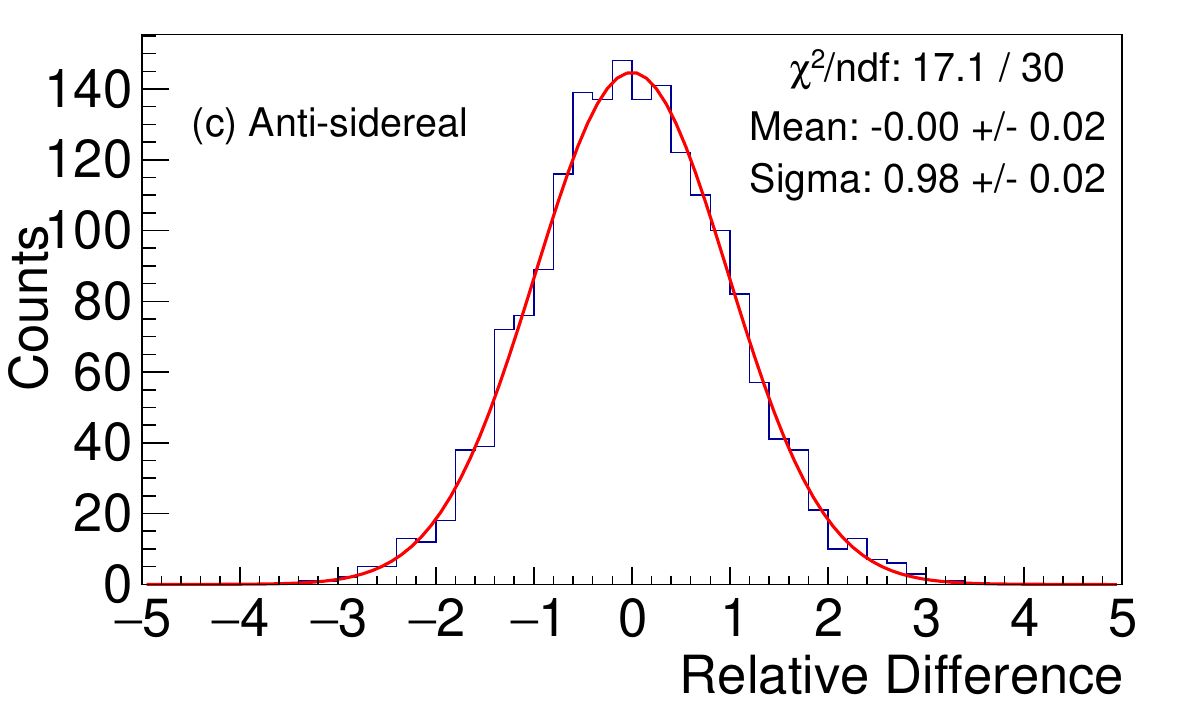}
  \end{minipage}%
  \hfill
  \begin{minipage}[b]{0.49\textwidth}
    \centering
    \includegraphics[width=\textwidth]{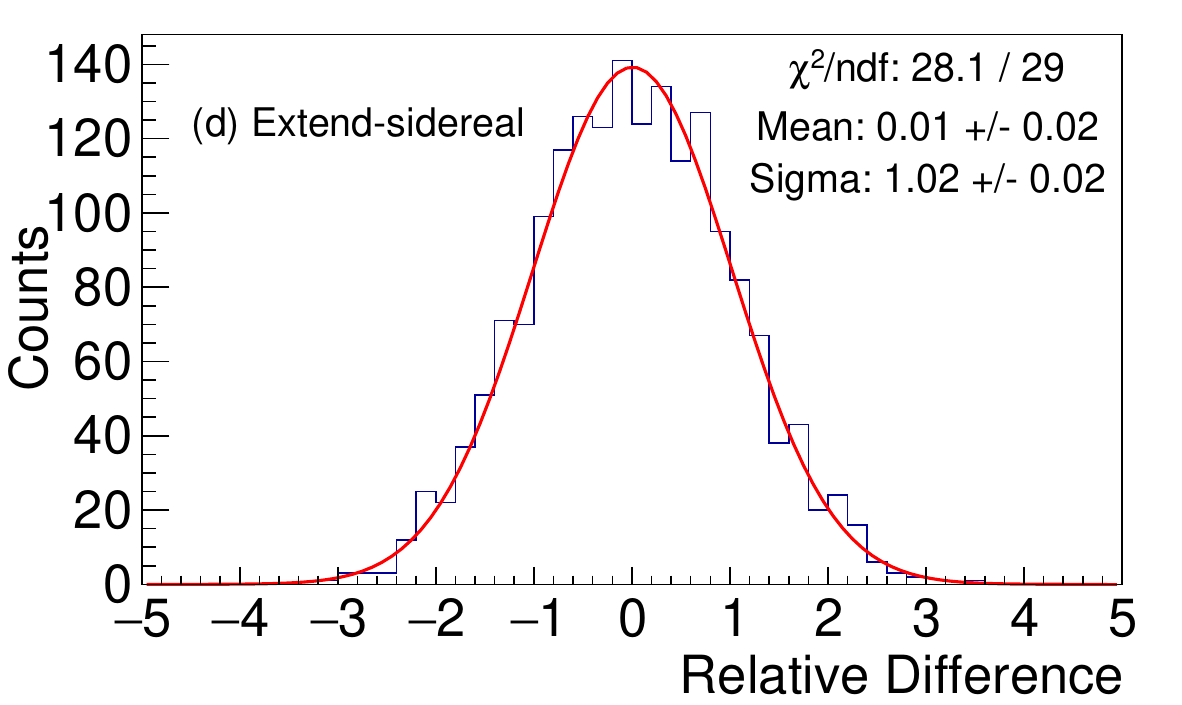}
  \end{minipage}

  \vspace{0.5em}
  \begin{minipage}[b]{0.49\textwidth}
    \centering
    \includegraphics[width=\textwidth]{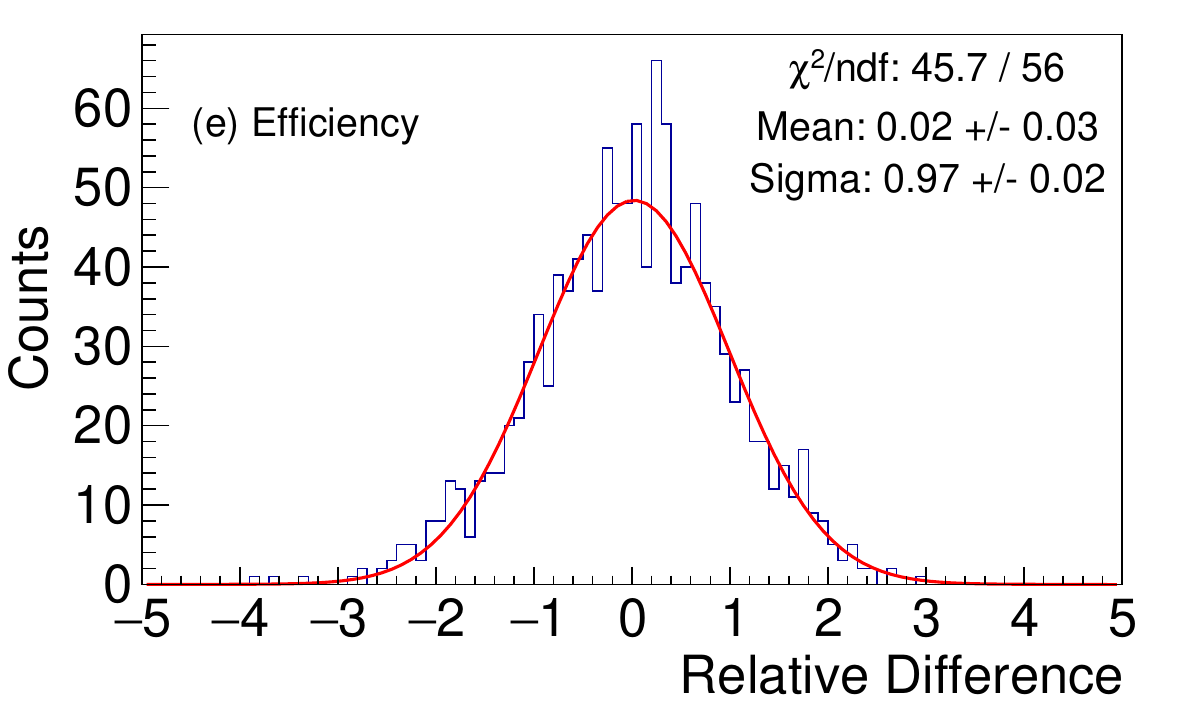}
  \end{minipage}

  \caption{The distributions of the relative difference for four time frames considering stable efficiency. Panels (a)-(d) correspond to solar, sidereal, anti-sidereal, and extended-sidereal time, respectively. Panel (e) shows the distribution of the relative difference with respect to the input efficiency for the case of stable efficiency. The red solid lines are the Gaussian fit. The fitted values of $\chi^2$, mean and standard deviation are indicated in the upper right corner of each panel.}
  \label{fig:4pull_dis_eff_stable}
\end{figure*}

\begin{figure*}[!htb]
\centering
\begin{minipage}{0.45\textwidth}
    \centering
    \includegraphics[width=\textwidth]{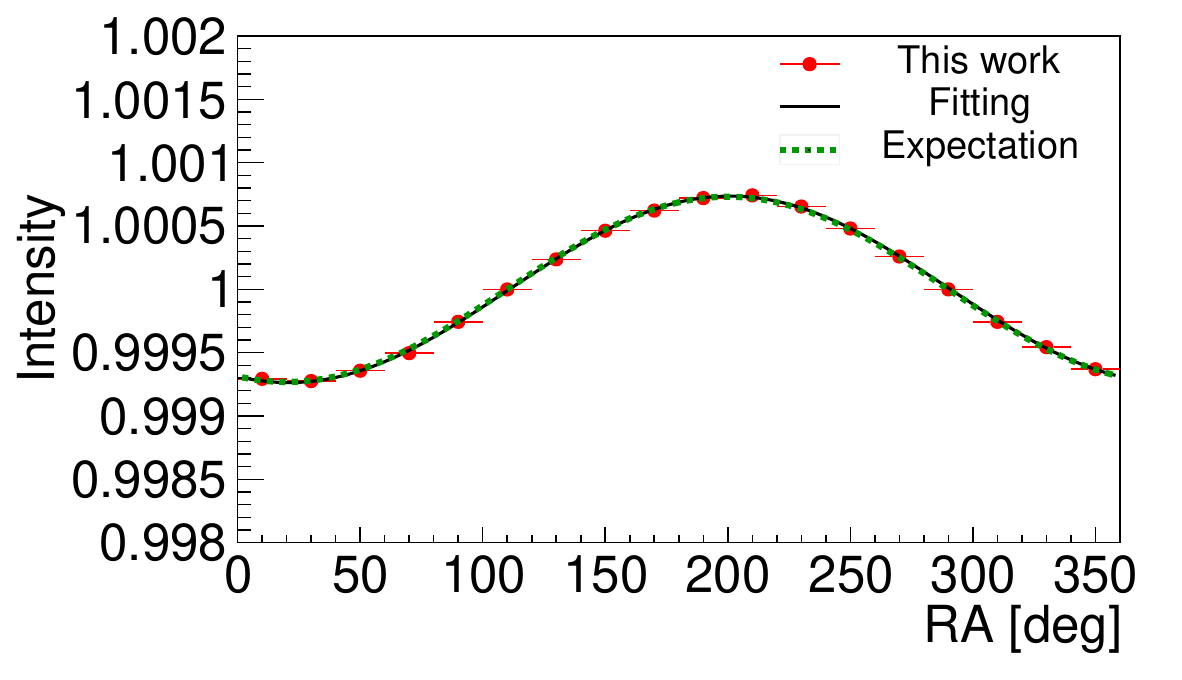}
\end{minipage}%
\hfill
\begin{minipage}{0.45\textwidth}
    \centering
    \includegraphics[width=\textwidth]{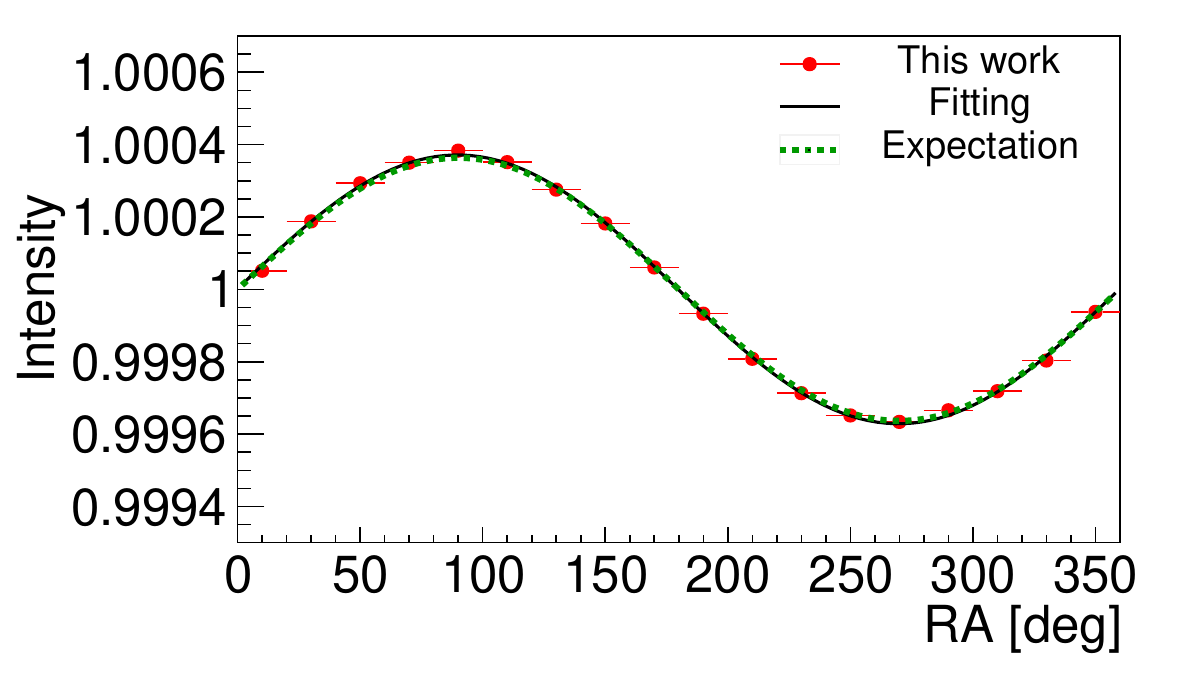}
\end{minipage}
\caption{
Comparison of the 1D distributions of relative intensity. The red dots and black solid line represent the 1D distributions obtained from the iteration and the corresponding fitting, respectively, while the green dotted line denotes the expected distribution. The left and right panels show the results for sidereal and solar time, respectively.
}
\label{fig:compare_1d}
\end{figure*}

To further quantitatively validate the iteration results, we compare the relative difference between the iterated map and the expected map on a grid-by-grid basis. The relative difference for each grid is defined as a standard normal variable, i.e.
\begin{equation}
\Delta I_{\rm iter}(\delta, \alpha) = \frac{I_{\rm iter}(\delta, \alpha) -I_{\rm exp}(\delta, \alpha)}{\sigma_{\rm iter}(\delta, \alpha)}
\label{eq:8}
\end{equation}
where $I_{\rm iter}$, $I_{\rm exp}$ and $\sigma_{\rm iter}$ represent the iterated intensity, expected intensity, and iterated uncertainty for each grid, respectively. If the differences arise solely from statistical fluctuations, the distribution of the relative differences of all grids is expected to follow a standard normal distribution, i.e. $N(0,1)$. Figure \ref{fig:4pull_dis_eff_stable} shows the distribution of the relative difference for the four time frames, each of which is approximately Gaussian. Gaussian fits are performed for all four distributions, and the resulting values of $\chi^2$, mean, and standard deviation are consistent with a standard normal distribution. This indicates that the iterated and expected maps agree within statistical fluctuations. \textcolor{black}{Following equation (\ref{eq:8}), we can also define the relative difference of efficiency in the local coordinate, allowing comparison between the iterated efficiency and the input efficiency:
\begin{equation}
\Delta \varepsilon_{\rm iter}(\theta, \phi) = \frac{\varepsilon_{\rm iter}(\theta, \phi) -\varepsilon_{\rm exp}(\theta, \phi)}{\sigma_{\rm iter}(\theta, \phi)}
\label{eq:8_2}
\end{equation}
Panel (e) of Fig. \ref{fig:4pull_dis_eff_stable} presents the distribution of relative difference of detection efficiency based on a grid-by-grid comparison. The iterated results are consistent with expectation within statistical fluctuations.
}

Figure (\ref{fig:compare_1d}) compares the one-dimensional profiles of relative intensity along right ascension between the iteration and expectation. This is frequently used when deriving the energy dependence of amplitude and phase of large-scale anisotropy. Each one-dimensional profile is obtained by averaging the relative intensities over all declination bands of the anisotropy map prior to smoothing. Here, right ascension is binned into $18$ intervals. The one-dimensional distribution is fitted by a first-order harmonic function, i.e.
\begin{align}
R(\alpha) = 1 + A_1 \cos(\alpha -\phi_1) ~,
\end{align}
where $A_1$ is the amplitude of the first harmonic, and $\phi_1$ is the phase at which $R(\alpha)$ reaches its maximum. The two panels show the comparisons for sidereal (left) and solar (right) time sky maps, respectively.

\subsection{Simultaneous fitting with unstable detection efficiency}

In the actual operation, the electronic components of detectors inevitably age or break down after long-term use, and the conditions of the detected medium within the detectors also change over time. For water Cherenkov detector arrays (WCDA) at LHAASO, for example, water quality deteriorates due to microbial contamination, necessitating frequent water purification to maintain stable operation of the array. In such scenarios, the assumption of stable efficiency is difficult to satisfy throughout the entire period, and detection efficiency may vary with time. 

If we can identify each time interval during which the detector array operates under stable conditions by studying the azimuthal distribution, the detection efficiency for each interval can also be determined by the enhanced method. In this way, the anisotropies can be reliably extracted without being affected by unstable operating conditions of the array. To mimic a time-varying detection efficiency, the response is modeled as a piecewise function of the following form:
\begin{equation}
\varepsilon(\theta, \phi, t) =
\begin{cases}
\begin{aligned}
& \cos^{8.5} \theta \, \sin \theta \, 
  \exp \Big[ -\dfrac{7.6}{\cos \theta} \Big] \\
& \quad \times [1+0.1 \sin(\phi)]~, 
  \quad 1 \leqslant t \leqslant 182
\end{aligned} \\
\begin{aligned}
& \sin \theta \, 
  \exp \Big[\dfrac{42.56}{1-\cos \theta} \Big] \\
& \quad \times [1+0.1 \cos(\phi)]~, 
  \quad 183 \leqslant t \leqslant 365
\end{aligned}
\end{cases}
\label{eq:effi_unstable}
\end{equation}
\begin{figure*}[!htb]
  \centering

  \begin{minipage}[b]{0.49\textwidth}
    \centering
    \includegraphics[width=\textwidth]{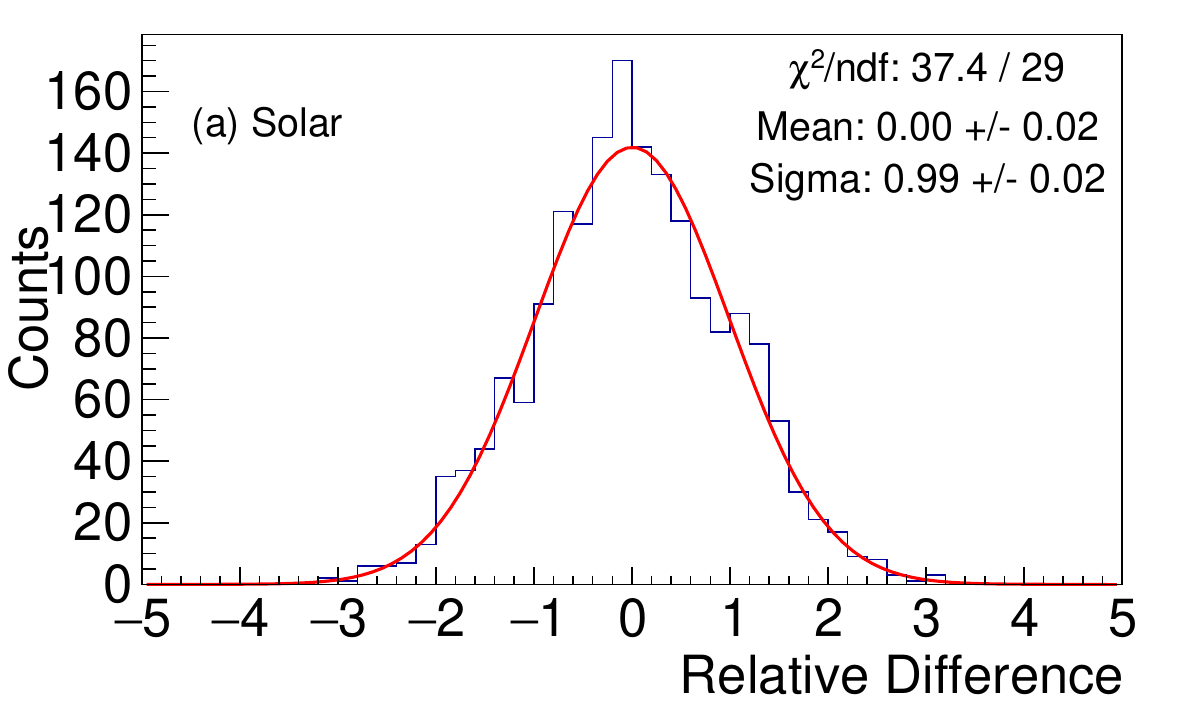}
  \end{minipage}%
  \hfill
  \begin{minipage}[b]{0.49\textwidth}
    \centering
    \includegraphics[width=\textwidth]{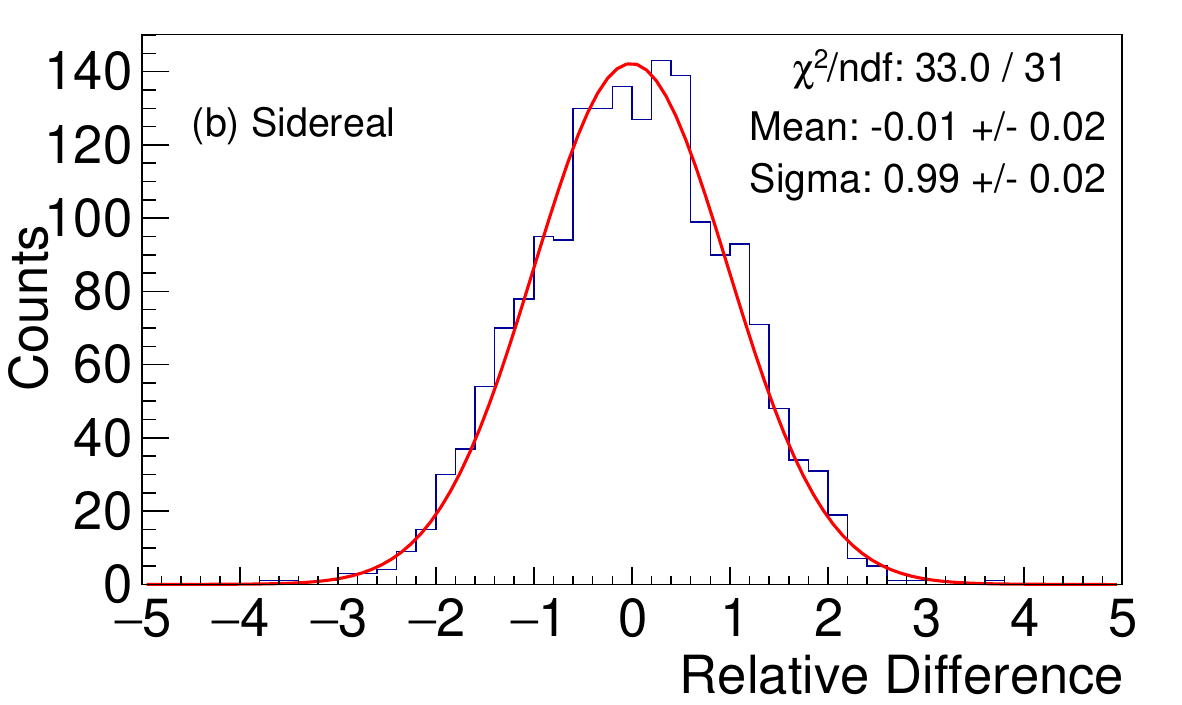}
  \end{minipage}

  \vspace{0.5em}
  \begin{minipage}[b]{0.49\textwidth}
    \centering
    \includegraphics[width=\textwidth]{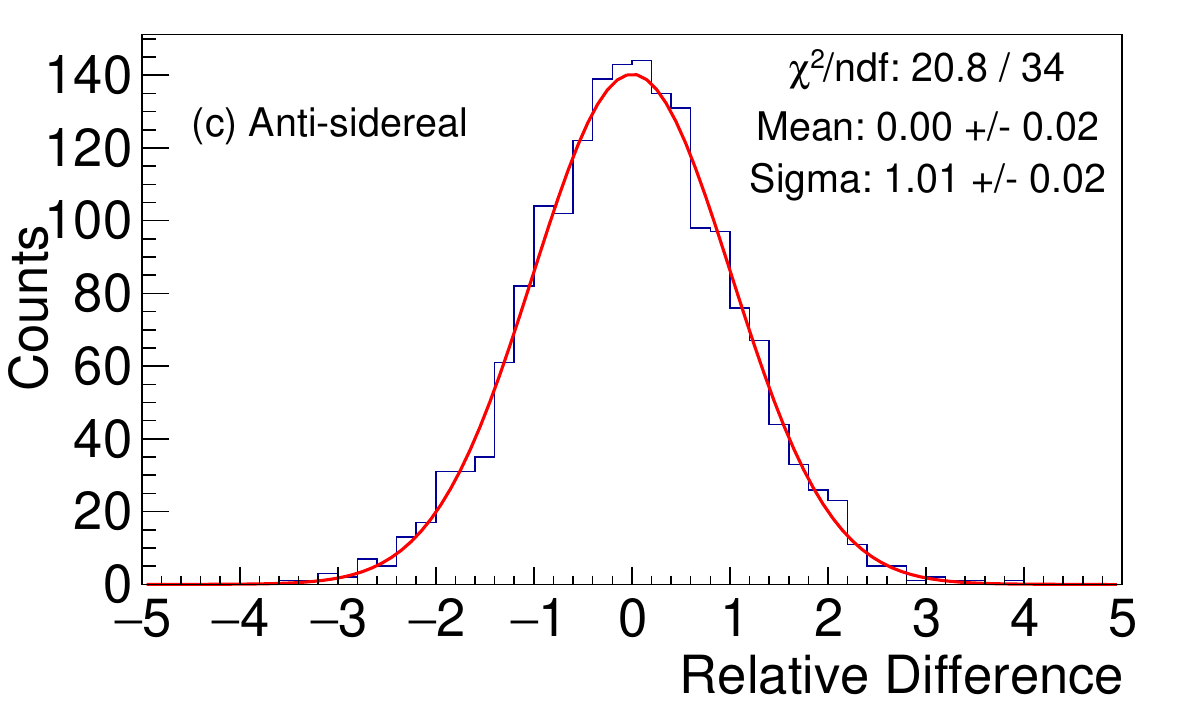}
  \end{minipage}%
  \hfill
  \begin{minipage}[b]{0.49\textwidth}
    \centering
    \includegraphics[width=\textwidth]{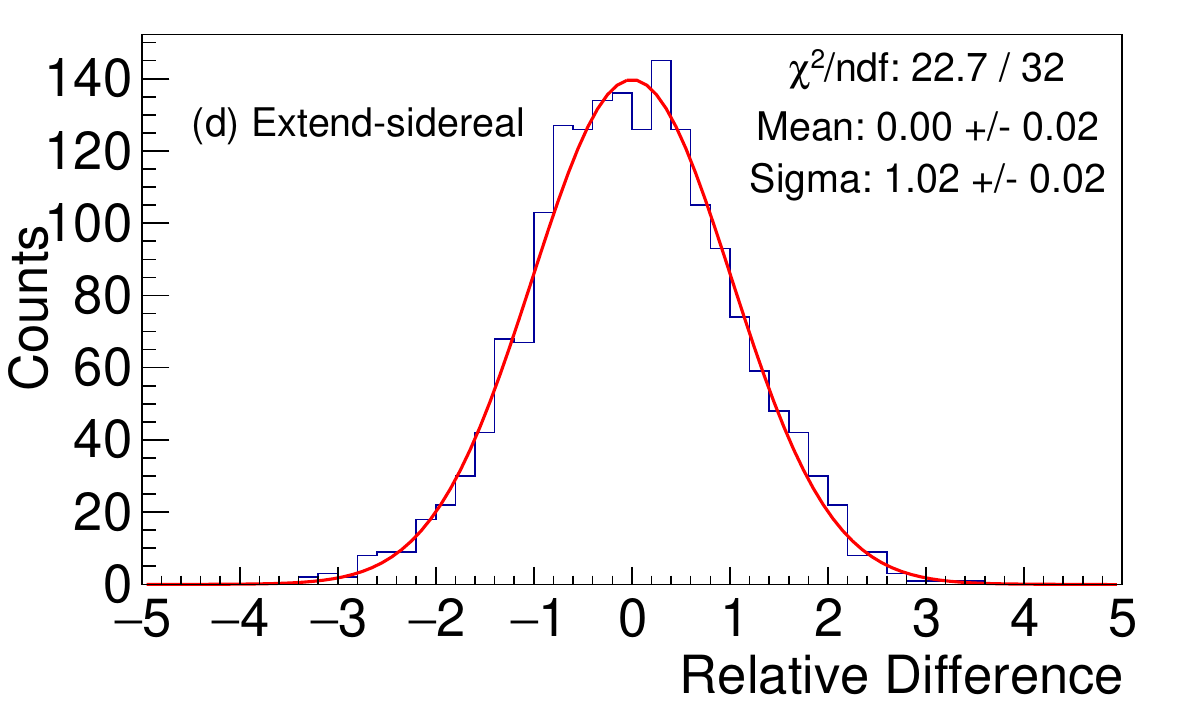}
  \end{minipage}

  \vspace{0.5em}
  \begin{minipage}[b]{0.49\textwidth}
    \centering
    \includegraphics[width=\textwidth]{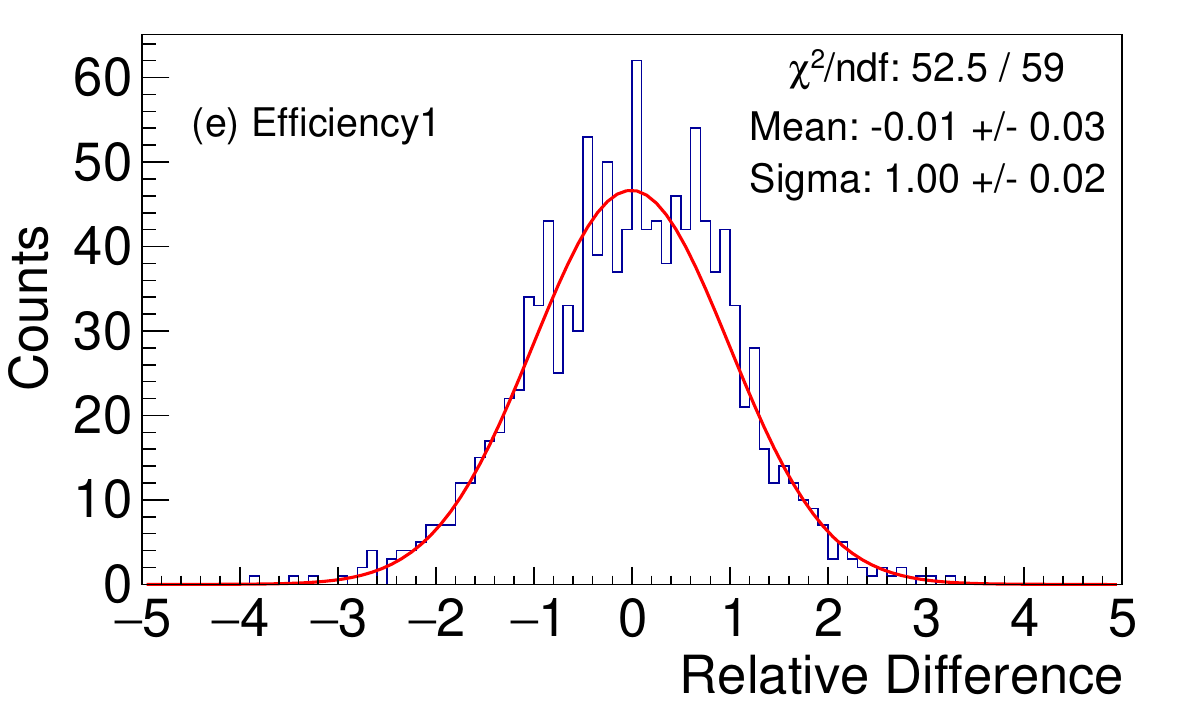}
  \end{minipage}%
  \hfill
  \begin{minipage}[b]{0.49\textwidth}
    \centering
    \includegraphics[width=\textwidth]{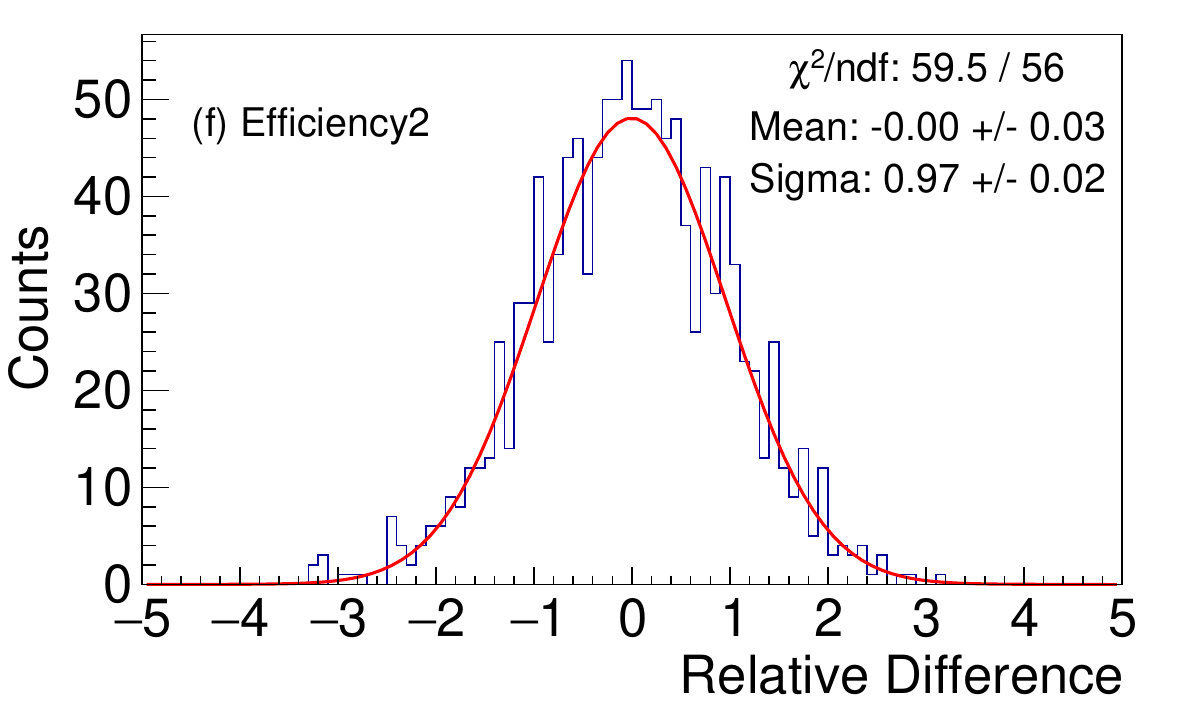}
  \end{minipage}

  \caption{The same distributions as Figure~\ref{fig:4pull_dis_eff_stable} but considering unstable detection efficiency. Panel (e) shows the distribution of the relative difference with respect to the input efficiency in the first time interval, and panel (f) presents the corresponding results for the second time interval.}
  \label{fig:4pull_dis_eff_unstable}
\end{figure*}

Figure (\ref{fig:4pull_dis_eff_unstable}) presents the distribution of relative difference between the iterated and expected sky maps based on a grid-by-grid comparison. All four reconstructed sky maps are found to be consistent with the expected ones, demonstrating that the enhanced method is capable of deriving anisotropies reliably even under unstable detector operating conditions. \textcolor{black}{For the unstable efficiency case, panel (e) in Figure (\ref{fig:4pull_dis_eff_unstable}) presents the relative difference between the iterated and input efficiency in the first 182 days, while panel (f) presents the relative difference of efficiency for the remaining days. The detection efficiencies for the case of unstable can be derived respectively.}

\subsection{Simultaneous fitting with incomplete tropical years}

\begin{figure*}[!htb]
\centering
\begin{minipage}{0.45\textwidth}
    \centering
    \includegraphics[width=\textwidth]{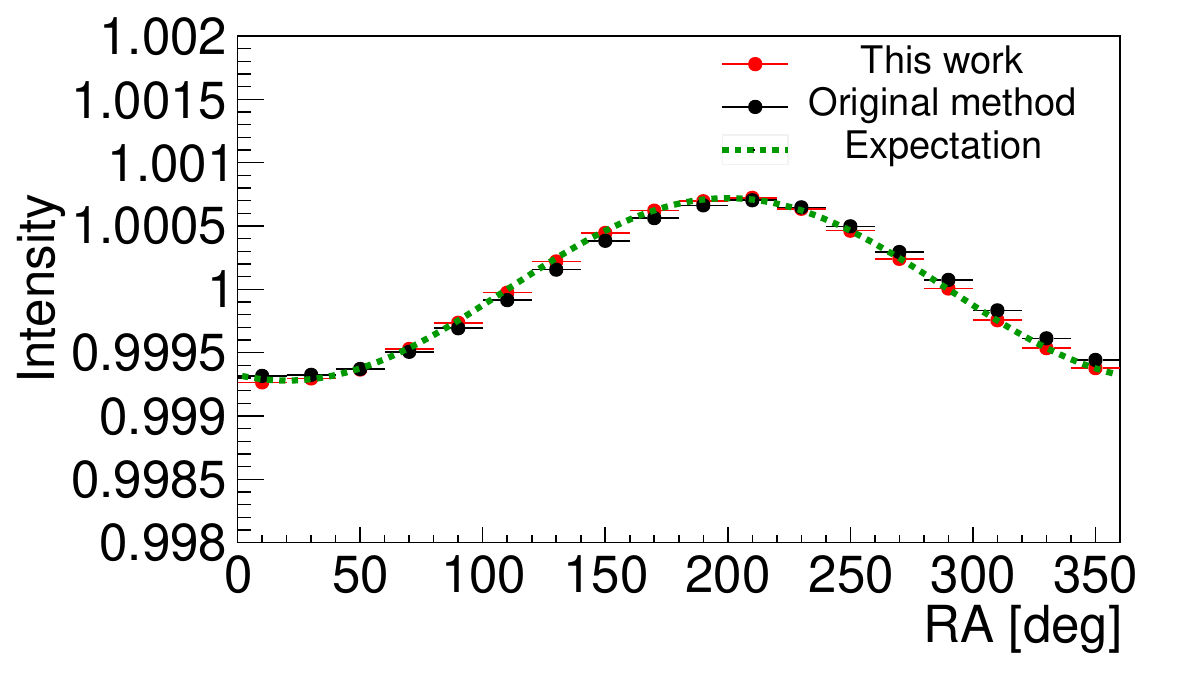}
    \label{fig:compare_500_sid}
\end{minipage}%
\hfill
\begin{minipage}{0.45\textwidth}
    \centering
    \includegraphics[width=\textwidth]{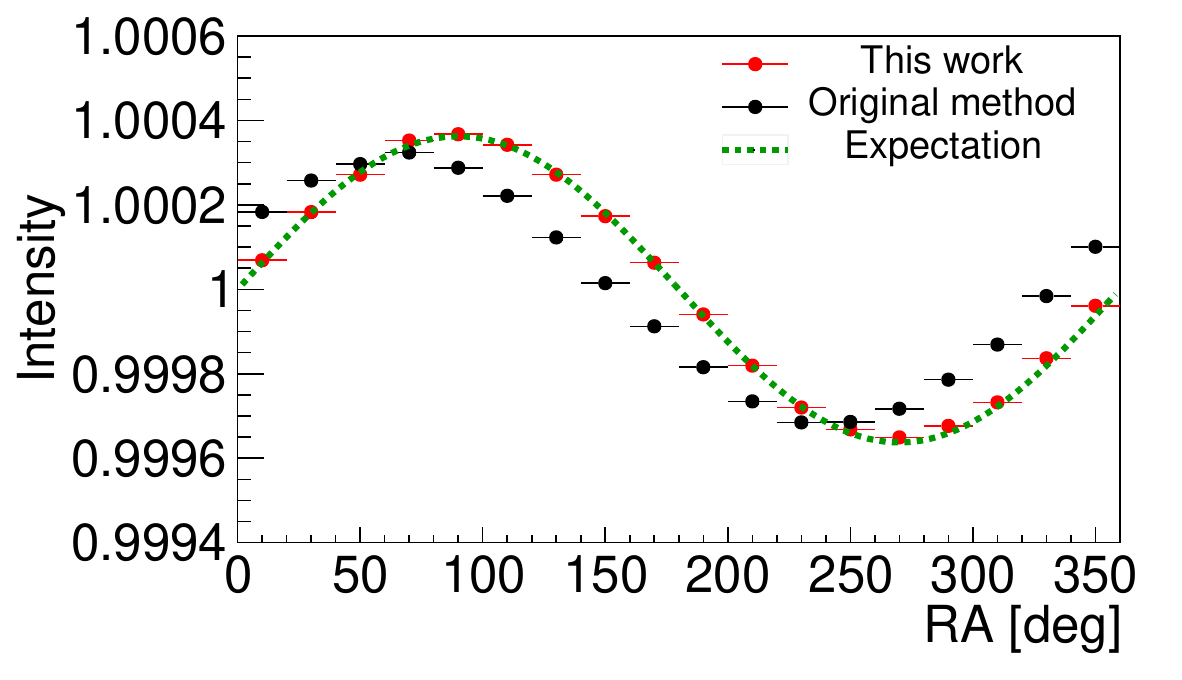}
    \label{fig:compare_500_solar}
\end{minipage}
\caption{
Comparison of the 1D relative intensity distributions obtained with the original (black dots) and enhanced (red dots) methods, for the case where the data set does not cover an integer number of tropical years. The green dashed line represents the expected distribution.
}
\label{fig:compare_500sd}
\end{figure*}

\begin{figure*}[!htb]
  \centering

  \begin{minipage}[b]{0.49\textwidth}
    \centering
    \includegraphics[width=\textwidth]{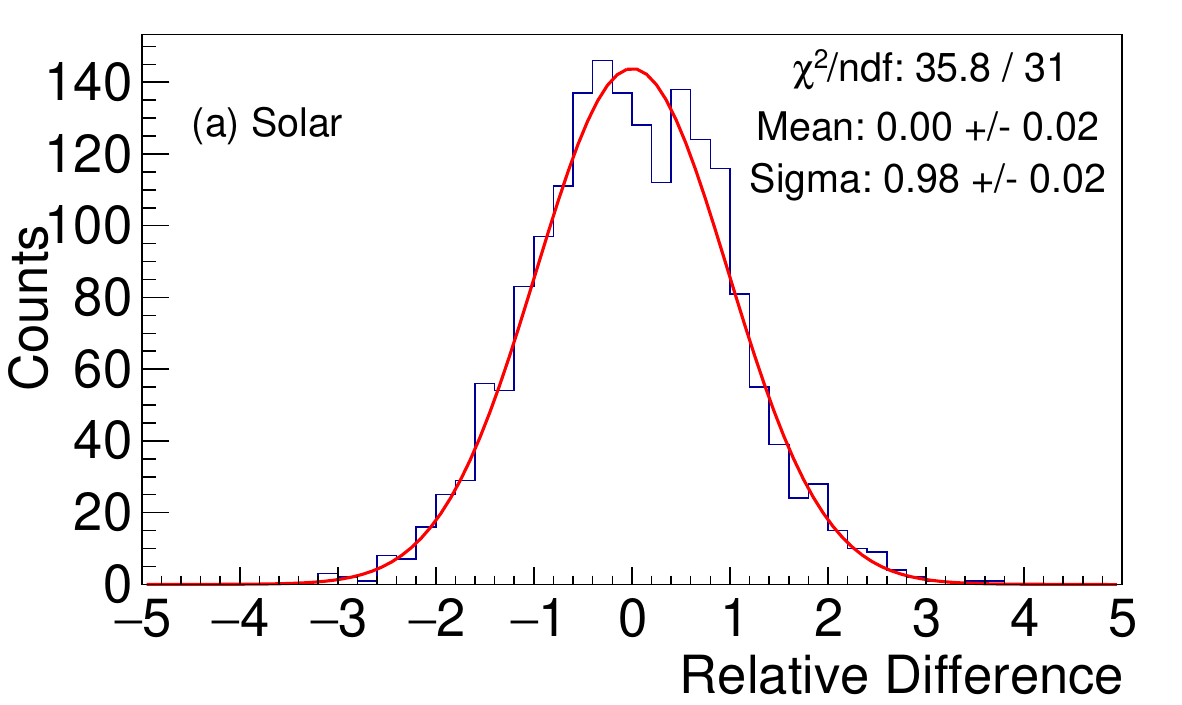}
  \end{minipage}%
  \hfill
  \begin{minipage}[b]{0.49\textwidth}
    \centering
    \includegraphics[width=\textwidth]{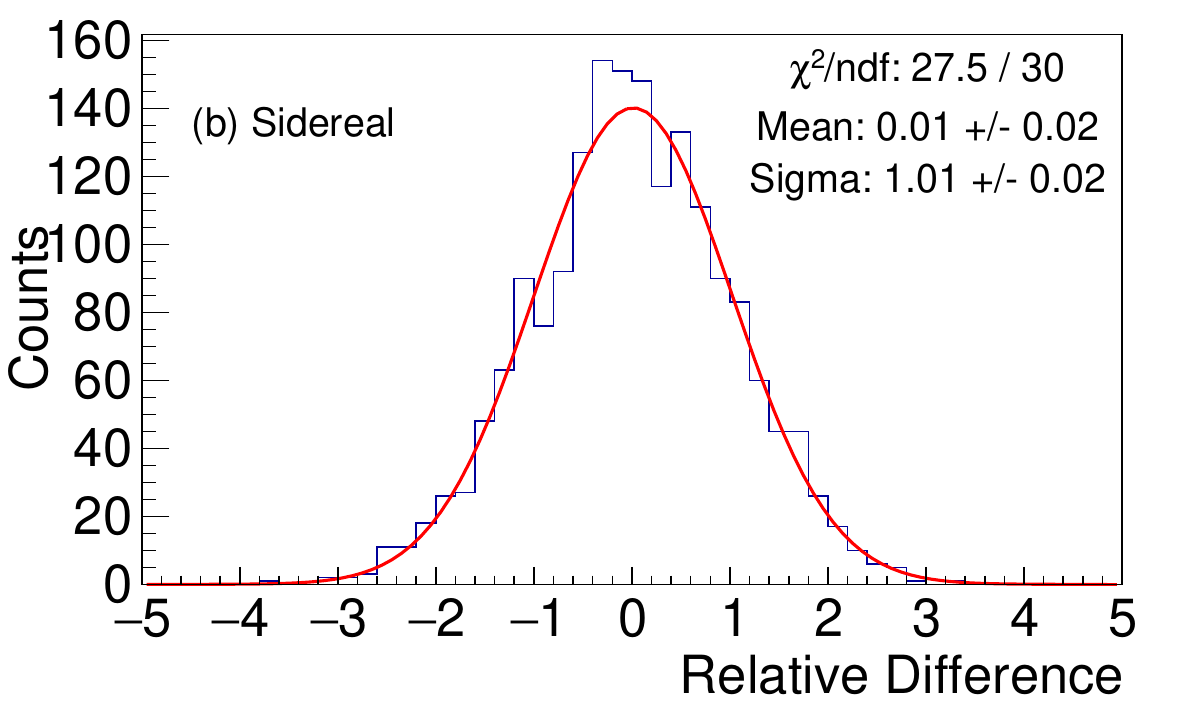}
  \end{minipage}

  \vspace{0.5em}
  \begin{minipage}[b]{0.49\textwidth}
    \centering
    \includegraphics[width=\textwidth]{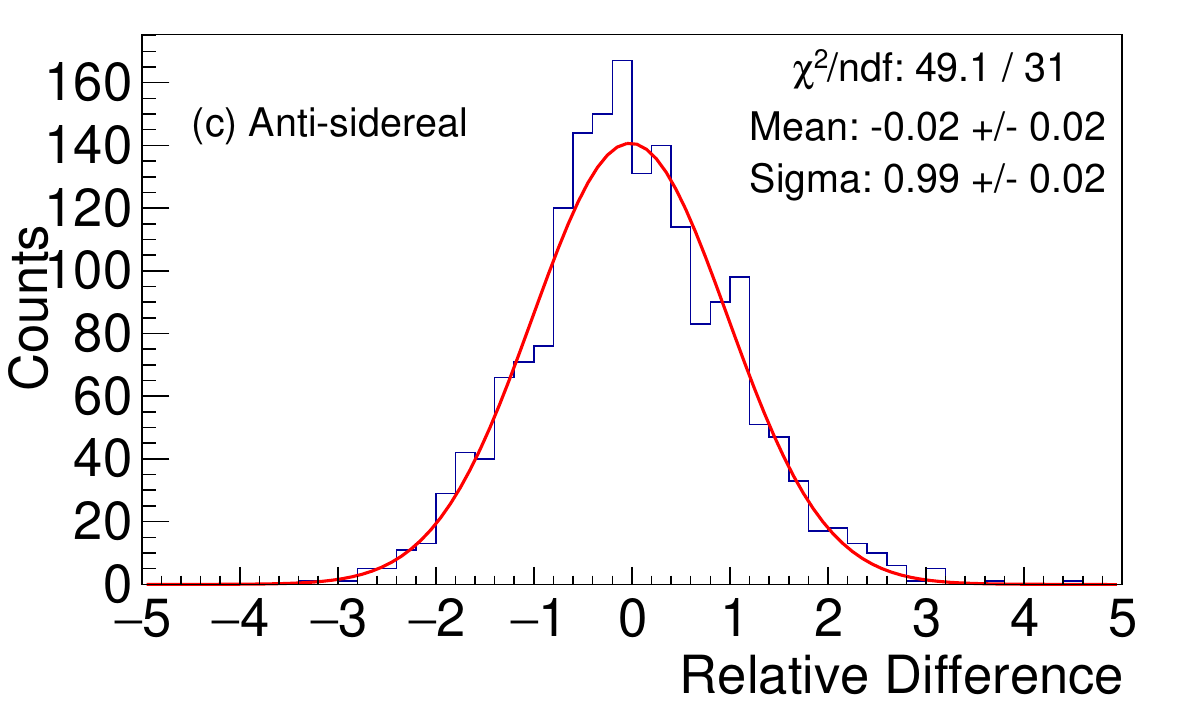}
  \end{minipage}%
  \hfill
  \begin{minipage}[b]{0.49\textwidth}
    \centering
    \includegraphics[width=\textwidth]{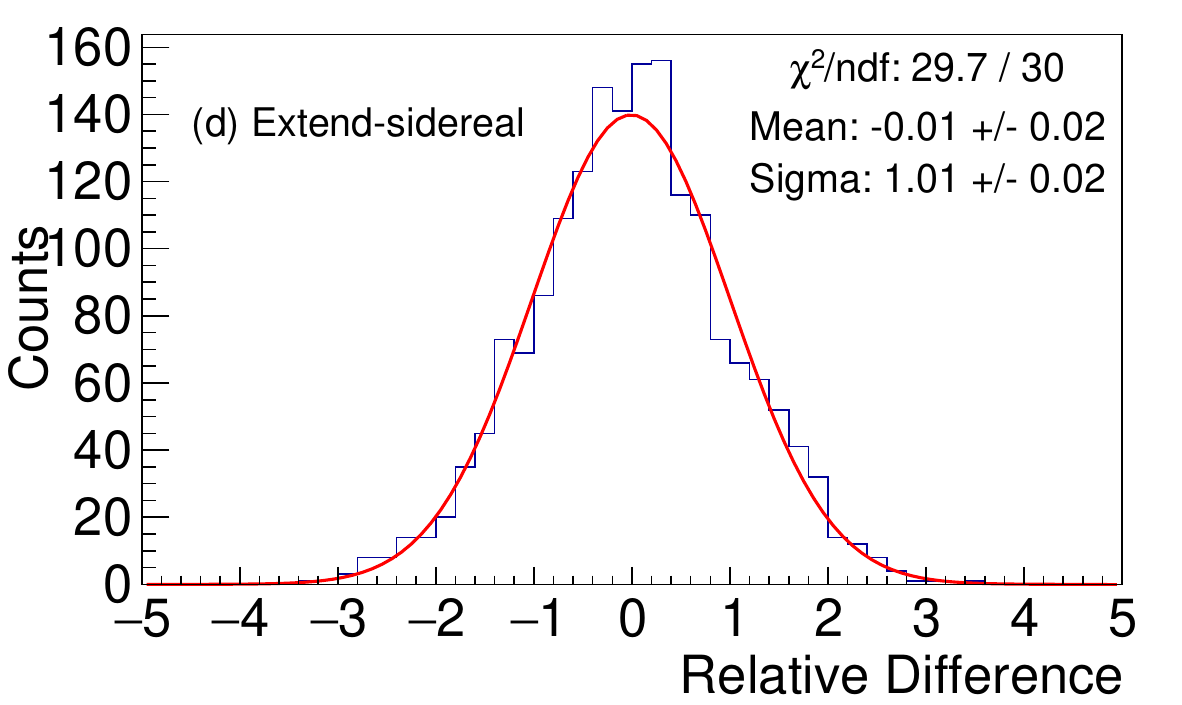}
  \end{minipage}

  \caption{The same distributions as Figure~\ref{fig:4pull_dis_eff_stable} but considering incomplete tropical years ($500$ solar days).}
  \label{fig:4pull_dis_eff_500sd}
\end{figure*}

We further illustrate the iteration results when the dataset does not cover an integer number of tropical years. As long as the observation time exceeds one tropical year—though not necessarily a full number of years—the sidereal and solar anisotropies can still be disentangled using the enhanced all-distance equi-zenith angle method. Figure \ref{fig:compare_500sd} compares the one-dimensional relative intensity profiles along right ascension derived by the original (black dots) and enhanced (red dots) methods, together with the expected profile (green dotted line). The left and right panels show the results for sidereal and solar time, respectively. In this case, a total observation time of $500$ solar days is assumed. Because the observation period does not span an integer number of tropical years, the reconstructed sidereal and solar profiles from the original method deviate from expectations, particularly for the solar anisotropy. In contrast, the results from the enhanced method remain consistent with the expected values. Figure \ref{fig:4pull_dis_eff_500sd} further presents the distribution of relative differences for the four time frames. Even with incomplete tropical years of observation, the iterated four anisotropy sky maps are also in accordance with expectations, except for statistical fluctuations.

\section{Summary and disscussion}

Due to the small amplitude of CR anisotropies, their detection has long been a challenging task. Instrumental and atmospheric artifacts can introduce spurious variations that interfere with the measurement of genuine anisotropy signals. Therefore, the elimination of instrumental and atmospheric effects must be carried out prior to analysis. 

The all-distance equi-zenith angle method was developed to reconstruct CR anisotropies by estimating the background using sideband data from the same zenith angle band observed simultaneously. \textcolor{black}{In its original formulation, however, the method requires long-term stability of the detector efficiency. Otherwise, the instability in the detector efficiency could affect the anisotropy measurements and a more careful correction of the azimuthal efficiency becomes necessary.} Moreover, the method is only applicable to datasets spanning complete tropical years, potentially leading to data wastage.

In this paper, we introduce an enhanced all-distance equi-zenith angle method. Unlike the original approach, the improved method allows for the simultaneous measurement of anisotropies across multiple time frames and the determination of the detector array's azimuthal efficiency without the need for prior correction of instrumental or atmospheric effects. This makes it particularly well suited for detector arrays operating under non-stationary conditions. Furthermore, the method no longer requires data to cover an integer number of tropical years. We evaluate the performance of the enhanced method using Monte Carlo simulations, and all of anisotropies at four time frames are found to be consistent with expectations.

Due to the scanning mode of ground-based observatories, which relies on Earth's rotation, measurements are largely insensitive to variations in CR anisotropy at declination bands. Therefore, for the dipole anisotropy, only its projection onto the celestial equator can be detected. Furthermore, in the current method, instrumental and atmospheric effects are degenerate and treated collectively as part of the detector efficiency. One solution to these limitations is the use of satellites, whose scanning strategy enables full-sky reconstruction of the anisotropies without projection effects. For ground-based observatories, an additional rotational degree of freedom would need to be introduced in order to measure anisotropies across declination bands. Then, instrumental and atmospheric effects could be determined separately through the iterative analysis. These aspects will be explored in future work.

\section*{Acknowledgements}
This work is supported by the National Natural Science Foundation of China (No. 12220101003, 12575120), the Project for Young Scientists in Basic Research of Chinese Academy of Sciences (No. YSBR-061), and National SKA Program of China grant (No. 2025SKA0110103).

\bibliographystyle{apsrev4-1}  
\bibliography{refs}           

@ARTICLE{2017PrPNP..94..184A,
       author = {{Ahlers}, Markus and {Mertsch}, Philipp},
        title = "{Origin of small-scale anisotropies in Galactic cosmic rays}",
      journal = {Progress in Particle and Nuclear Physics},
         year = 2017,
        month = may,
       volume = {94},
        pages = {184-216},
          doi = {10.1016/j.ppnp.2017.01.004},
archivePrefix = {arXiv},
       eprint = {1612.01873},
 primaryClass = {astro-ph.HE},
       adsurl = {https://ui.adsabs.harvard.edu/abs/2017PrPNP..94..184A}
}

@ARTICLE{2003ApJ...595..803A,
       author = {{Atkins}, R. and {Benbow}, W. and {Berley}, D. and {Blaufuss}, E. and {Bussons}, J. and {Coyne}, D.~G. and {Delay}, R.~S. and {De Young}, T. and {Dingus}, B.~L. and {Dorfan}, D.~E. and {Ellsworth}, R.~W. and {Falcone}, A. and {Fleysher}, L. and {Fleysher}, R. and {Gisler}, G. and {Gonzalez}, M.~M. and {Goodman}, J.~A. and {Haines}, T.~J. and {Hays}, E. and {Hoffman}, C.~M. and {Kelley}, L.~A. and {Laird}, R.~W. and {McCullough}, J. and {McEnery}, J.~E. and {Miller}, R.~S. and {Mincer}, A.~I. and {Morales}, M.~F. and {Nemethy}, P. and {Noyes}, D. and {Ryan}, J.~M. and {Samuelson}, F.~W. and {Schneider}, M. and {Shen}, B. and {Shoup}, A. and {Sinnis}, G. and {Smith}, A.~J. and {Sullivan}, G.~W. and {Tumer}, O.~T. and {Wang}, K. and {Wascko}, M. and {Williams}, D.~A. and {Westerhoff}, S. and {Wilson}, M.~E. and {Xu}, X. and {Yodh}, G.~B.},
        title = "{Observation of TeV Gamma Rays from the Crab Nebula with Milagro Using a New Background Rejection Technique}",
      journal = {\apj},
         year = 2003,
        month = oct,
       volume = {595},
       number = {2},
        pages = {803-811},
          doi = {10.1086/377498},
archivePrefix = {arXiv},
       eprint = {astro-ph/0305308},
 primaryClass = {astro-ph},
       adsurl = {https://ui.adsabs.harvard.edu/abs/2003ApJ...595..803A}
}

@ARTICLE{2005ApJ...633.1005A,
       author = {{Amenomori}, M. and {Ayabe}, S. and {Chen}, D. and {Cui}, S.~W. and {Danzengluobu} and {Ding}, L.~K. and {Ding}, X.~H. and {Feng}, C.~F. and {Feng}, Z.~Y. and {Gao}, X.~Y. and {Geng}, Q.~X. and {Guo}, H.~W. and {He}, H.~H. and {He}, M. and {Hibino}, K. and {Hotta}, N. and {Hu}, Haibing and {Hu}, H.~B. and {Huang}, J. and {Huang}, Q. and {Jia}, H.~Y. and {Kajino}, F. and {Kasahara}, K. and {Katayose}, Y. and {Kato}, C. and {Kawata}, K. and {Labaciren} and {Le}, G.~M. and {Li}, J.~Y. and {Lu}, H. and {Lu}, S.~L. and {Meng}, X.~R. and {Mizutani}, K. and {Mu}, J. and {Munakata}, K. and {Nagai}, A. and {Nanjo}, H. and {Nishizawa}, M. and {Ohnishi}, M. and {Ohta}, I. and {Onuma}, H. and {Ouchi}, T. and {Ozawa}, S. and {Ren}, J.~R. and {Saito}, T. and {Sakata}, M. and {Sasaki}, T. and {Shibata}, M. and {Shiomi}, A. and {Shirai}, T. and {Sugimoto}, H. and {Takashima}, M. and {Takita}, M. and {Tan}, Y.~H. and {Tateyama}, N. and {Torii}, S. and {Tsuchiya}, H. and {Udo}, S. and {Utsugi}, T. and {Wang}, H. and {Wang}, X. and {Wang}, Y.~G. and {Wu}, H.~R. and {Xue}, L. and {Yamamoto}, Y. and {Yan}, C.~T. and {Yang}, X.~C. and {Yasue}, S. and {Ye}, Z.~H. and {Yu}, G.~C. and {Yuan}, A.~F. and {Yuda}, T. and {Zhang}, H.~M. and {Zhang}, J.~L. and {Zhang}, N.~J. and {Zhang}, X.~Y. and {Zhang}, Yi and {Zhang}, Y. and {Zhaxisangzhu} and {Zhou}, X.~X. and {Tibet As{\ensuremath{\gamma}} Collaboration}},
        title = "{A Northern Sky Survey for Steady Tera-Electron Volt Gamma-Ray Point Sources Using the Tibet Air Shower Array}",
      journal = {\apj},
         year = 2005,
        month = nov,
       volume = {633},
       number = {2},
        pages = {1005-1012},
          doi = {10.1086/491612},
archivePrefix = {arXiv},
       eprint = {astro-ph/0502039},
 primaryClass = {astro-ph},
       adsurl = {https://ui.adsabs.harvard.edu/abs/2005ApJ...633.1005A}
}

@ARTICLE{1993NIMPA.328..570A,
       author = {{Alexandreas}, D.~E. and {Berley}, D. and {Biller}, S. and {Dion}, G.~M. and {Goodman}, J.~A. and {Haines}, T.~J. and {Hoffman}, C.~M. and {Horch}, E. and {Lu}, X.-Q. and {Sinnis}, C. and {Yodh}, G.~B. and {Zhang}, W.},
        title = "{Point source search techniques in ultra high energy gamma ray astronomy}",
      journal = {Nuclear Instruments and Methods in Physics Research A},
         year = 1993,
        month = may,
       volume = {328},
       number = {3},
        pages = {570-577},
          doi = {10.1016/0168-9002(93)90677-A},
       adsurl = {https://ui.adsabs.harvard.edu/abs/1993NIMPA.328..570A}
}

@ARTICLE{2006Sci...314..439A,
       author = {{Amenomori}, M. and {Ayabe}, S. and {Bi}, X.~J. and {Chen}, D. and {Cui}, S.~W. and {Danzengluobu} and {Ding}, L.~K. and {Ding}, X.~H. and {Feng}, C.~F. and {Feng}, Zhaoyang and {Feng}, Z.~Y. and {Gao}, X.~Y. and {Geng}, Q.~X. and {Guo}, H.~W. and {He}, H.~H. and {He}, M. and {Hibino}, K. and {Hotta}, N. and {Hu}, Haibing and {Hu}, H.~B. and {Huang}, J. and {Huang}, Q. and {Jia}, H.~Y. and {Kajino}, F. and {Kasahara}, K. and {Katayose}, Y. and {Kato}, C. and {Kawata}, K. and {Labaciren} and {Le}, G.~M. and {Li}, A.~F. and {Li}, J.~Y. and {Lou}, Y.-Q. and {Lu}, H. and {Lu}, S.~L. and {Meng}, X.~R. and {Mizutani}, K. and {Mu}, J. and {Munakata}, K. and {Nagai}, A. and {Nanjo}, H. and {Nishizawa}, M. and {Ohnishi}, M. and {Ohta}, I. and {Onuma}, H. and {Ouchi}, T. and {Ozawa}, S. and {Ren}, J.~R. and {Saito}, T. and {Saito}, T.~Y. and {Sakata}, M. and {Sako}, T.~K. and {Sasaki}, T. and {Shibata}, M. and {Shiomi}, A. and {Shirai}, T. and {Sugimoto}, H. and {Takita}, M. and {Tan}, Y.~H. and {Tateyama}, N. and {Torii}, S. and {Tsuchiya}, H. and {Udo}, S. and {Wang}, B. and {Wang}, H. and {Wang}, X. and {Wang}, Y.~G. and {Wu}, H.~R. and {Xue}, L. and {Yamamoto}, Y. and {Yan}, C.~T. and {Yang}, X.~C. and {Yasue}, S. and {Ye}, Z.~H. and {Yu}, G.~C. and {Yuan}, A.~F. and {Yuda}, T. and {Zhang}, H.~M. and {Zhang}, J.~L. and {Zhang}, N.~J. and {Zhang}, X.~Y. and {Zhang}, Y. and {Zhang}, Yi and {Zhaxisangzhu} and {Zhou}, X.~X. and {Tibet AS{\ensuremath{\gamma}} Collaboration}},
        title = "{Anisotropy and Corotation of Galactic Cosmic Rays}",
      journal = {Science},
         year = 2006,
        month = oct,
       volume = {314},
       number = {5798},
        pages = {439-443},
          doi = {10.1126/science.1131702},
archivePrefix = {arXiv},
       eprint = {astro-ph/0610671},
 primaryClass = {astro-ph},
       adsurl = {https://ui.adsabs.harvard.edu/abs/2006Sci...314..439A}
}

@ARTICLE{2007PhRvD..75f2003G,
       author = {{Guillian}, G. and {Hosaka}, J. and {Ishihara}, K. and {Kameda}, J. and {Koshio}, Y. and {Minamino}, A. and {Mitsuda}, C. and {Miura}, M. and {Moriyama}, S. and {Nakahata}, M. and {Namba}, T. and {Obayashi}, Y. and {Ogawa}, H. and {Shiozawa}, M. and {Suzuki}, Y. and {Takeda}, A. and {Takeuchi}, Y. and {Yamada}, S. and {Higuchi}, I. and {Ishitsuka}, M. and {Kajita}, T. and {Kaneyuki}, K. and {Mitsuka}, G. and {Nakayama}, S. and {Nishino}, H. and {Okada}, A. and {Okumura}, K. and {Saji}, C. and {Takenaga}, Y. and {Desai}, S. and {Kearns}, E. and {Stone}, J.~L. and {Sulak}, L.~R. and {Wang}, W. and {Goldhaber}, M. and {Casper}, D. and {Gajewski}, W. and {Griskevich}, J. and {Kropp}, W.~R. and {Liu}, D.~W. and {Mine}, S. and {Smy}, M.~B. and {Sobel}, H.~W. and {Vagins}, M.~R. and {Ganezer}, K.~S. and {Hill}, J. and {Keig}, W.~E. and {Scholberg}, K. and {Walter}, C.~W. and {Ellsworth}, R.~W. and {Tasaka}, S. and {Kibayashi}, A. and {Learned}, J.~G. and {Matsuno}, S. and {Messier}, M.~D. and {Hayato}, Y. and {Ichikawa}, A.~K. and {Ishida}, T. and {Ishii}, T. and {Iwashita}, T. and {Kobayashi}, T. and {Nakadaira}, T. and {Nakamura}, K. and {Nitta}, K. and {Oyama}, Y. and {Totsuka}, Y. and {Suzuki}, A.~T. and {Hasegawa}, M. and {Kato}, I. and {Maesaka}, H. and {Nakaya}, T. and {Nishikawa}, K. and {Sato}, H. and {Yamamoto}, S. and {Yokoyama}, M. and {Haines}, T.~J. and {Dazeley}, S. and {Hatakeyama}, S. and {Svoboda}, R. and {Blaufuss}, E. and {Goodman}, J.~A. and {Sullivan}, G.~W. and {Turcan}, D. and {Habig}, A. and {Fukuda}, Y. and {Itow}, Y. and {Sakuda}, M. and {Yoshida}, M. and {Kim}, S.~B. and {Yoo}, J. and {Okazawa}, H. and {Ishizuka}, T. and {Jung}, C.~K. and {Kato}, T. and {Kobayashi}, K. and {Malek}, M. and {Mauger}, C. and {McGrew}, C. and {Sharkey}, E. and {Yanagisawa}, C. and {Gando}, Y. and {Hasegawa}, T. and {Inoue}, K. and {Shirai}, J. and {Suzuki}, A. and {Nishijima}, K. and {Ishino}, H. and {Watanabe}, Y. and {Koshiba}, M. and {Kielczewska}, D. and {Berns}, H.~G. and {Gran}, R. and {Shiraishi}, K.~K. and {Stachyra}, A.~L. and {Washburn}, K. and {Wilkes}, R.~J. and {Munakata}, K.},
        title = "{Observation of the anisotropy of 10TeV primary cosmic ray nuclei flux with the Super-Kamiokande-I detector}",
      journal = {\prd},
         year = 2007,
        month = mar,
       volume = {75},
       number = {6},
          eid = {062003},
        pages = {062003},
          doi = {10.1103/PhysRevD.75.062003},
archivePrefix = {arXiv},
       eprint = {astro-ph/0508468},
 primaryClass = {astro-ph},
       adsurl = {https://ui.adsabs.harvard.edu/abs/2007PhRvD..75f2003G}
}

@ARTICLE{2008PhRvL.101v1101A,
       author = {{Abdo}, A.~A. and {Allen}, B. and {Aune}, T. and {Berley}, D. and {Blaufuss}, E. and {Casanova}, S. and {Chen}, C. and {Dingus}, B.~L. and {Ellsworth}, R.~W. and {Fleysher}, L. and {Fleysher}, R. and {Gonzalez}, M.~M. and {Goodman}, J.~A. and {Hoffman}, C.~M. and {H{\"u}ntemeyer}, P.~H. and {Kolterman}, B.~E. and {Lansdell}, C.~P. and {Linnemann}, J.~T. and {McEnery}, J.~E. and {Mincer}, A.~I. and {Nemethy}, P. and {Noyes}, D. and {Pretz}, J. and {Ryan}, J.~M. and {Parkinson}, P.~M. Saz and {Shoup}, A. and {Sinnis}, G. and {Smith}, A.~J. and {Sullivan}, G.~W. and {Vasileiou}, V. and {Walker}, G.~P. and {Williams}, D.~A. and {Yodh}, G.~B.},
        title = "{Discovery of Localized Regions of Excess 10-TeV Cosmic Rays}",
      journal = {\prl},
         year = 2008,
        month = nov,
       volume = {101},
       number = {22},
          eid = {221101},
        pages = {221101},
          doi = {10.1103/PhysRevLett.101.221101},
archivePrefix = {arXiv},
       eprint = {0801.3827},
 primaryClass = {astro-ph},
       adsurl = {https://ui.adsabs.harvard.edu/abs/2008PhRvL.101v1101A}
}

@ARTICLE{2009ApJ...698.2121A,
       author = {{Abdo}, A.~A. and {Allen}, B.~T. and {Aune}, T. and {Berley}, D. and {Casanova}, S. and {Chen}, C. and {Dingus}, B.~L. and {Ellsworth}, R.~W. and {Fleysher}, L. and {Fleysher}, R. and {Gonzalez}, M.~M. and {Goodman}, J.~A. and {Hoffman}, C.~M. and {Hopper}, B. and {H{\"u}ntemeyer}, P.~H. and {Kolterman}, B.~E. and {Lansdell}, C.~P. and {Linnemann}, J.~T. and {McEnery}, J.~E. and {Mincer}, A.~I. and {Nemethy}, P. and {Noyes}, D. and {Pretz}, J. and {Ryan}, J.~M. and {Parkinson}, P.~M. Saz and {Shoup}, A. and {Sinnis}, G. and {Smith}, A.~J. and {Sullivan}, G.~W. and {Vasileiou}, V. and {Walker}, G.~P. and {Williams}, D.~A. and {Yodh}, G.~B.},
        title = "{The Large-Scale Cosmic-Ray Anisotropy as Observed with Milagro}",
      journal = {\apj},
         year = 2009,
        month = jun,
       volume = {698},
       number = {2},
        pages = {2121-2130},
          doi = {10.1088/0004-637X/698/2/2121},
archivePrefix = {arXiv},
       eprint = {0806.2293},
 primaryClass = {astro-ph},
       adsurl = {https://ui.adsabs.harvard.edu/abs/2009ApJ...698.2121A}
}

@ARTICLE{2010ApJ...718L.194A,
       author = {{Abbasi}, R. and {Abdou}, Y. and {Abu-Zayyad}, T. and {Adams}, J. and {Aguilar}, J.~A. and {Ahlers}, M. and {Andeen}, K. and {Auffenberg}, J. and {Bai}, X. and {Baker}, M. and {Barwick}, S.~W. and {Bay}, R. and {Bazo Alba}, J.~L. and {Beattie}, K. and {Beatty}, J.~J. and {Bechet}, S. and {Becker}, J.~K. and {Becker}, K.-H. and {Benabderrahmane}, M.~L. and {BenZvi}, S. and {Berdermann}, J. and {Berghaus}, P. and {Berley}, D. and {Bernardini}, E. and {Bertrand}, D. and {Besson}, D.~Z. and {Bissok}, M. and {Blaufuss}, E. and {Boersma}, D.~J. and {Bohm}, C. and {B{\"o}ser}, S. and {Botner}, O. and {Bradley}, L. and {Braun}, J. and {Buitink}, S. and {Carson}, M. and {Chirkin}, D. and {Christy}, B. and {Clem}, J. and {Clevermann}, F. and {Cohen}, S. and {Colnard}, C. and {Cowen}, D.~F. and {D'Agostino}, M.~V. and {Danninger}, M. and {Davis}, J.~C. and {De Clercq}, C. and {Demir{\"o}rs}, L. and {Depaepe}, O. and {Descamps}, F. and {Desiati}, P. and {de Vries-Uiterweerd}, G. and {DeYoung}, T. and {D{\'\i}az-V{\'e}lez}, J.~C. and {Dierckxsens}, M. and {Dreyer}, J. and {Dumm}, J.~P. and {Duvoort}, M.~R. and {Ehrlich}, R. and {Eisch}, J. and {Ellsworth}, R.~W. and {Engdeg{\r{a}}rd}, O. and {Euler}, S. and {Evenson}, P.~A. and {Fadiran}, O. and {Fazely}, A.~R. and {Feusels}, T. and {Filimonov}, K. and {Finley}, C. and {Foerster}, M.~M. and {Fox}, B.~D. and {Franckowiak}, A. and {Franke}, R. and {Gaisser}, T.~K. and {Gallagher}, J. and {Geisler}, M. and {Gerhardt}, L. and {Gladstone}, L. and {Gl{\"u}senkamp}, T. and {Goldschmidt}, A. and {Goodman}, J.~A. and {Grant}, D. and {Griesel}, T. and {Gro{\ss}}, A. and {Grullon}, S. and {Gurtner}, M. and {Ha}, C. and {Hallgren}, A. and {Halzen}, F. and {Han}, K. and {Hanson}, K. and {Helbing}, K. and {Herquet}, P. and {Hickford}, S. and {Hill}, G.~C. and {Hoffman}, K.~D. and {Homeier}, A. and {Hoshina}, K. and {Hubert}, D. and {Huelsnitz}, W. and {H{\"u}l{\ss}}, J.-P. and {Hulth}, P.~O. and {Hultqvist}, K. and {Hussain}, S. and {Ishihara}, A. and {Jacobsen}, J. and {Japaridze}, G.~S. and {Johansson}, H. and {Joseph}, J.~M. and {Kampert}, K.-H. and {Karg}, T. and {Karle}, A. and {Kelley}, J.~L. and {Kemming}, N. and {Kenny}, P. and {Kiryluk}, J. and {Kislat}, F. and {Klein}, S.~R. and {Knops}, S. and {K{\"o}hne}, J.-H. and {Kohnen}, G. and {Kolanoski}, H. and {K{\"o}pke}, L. and {Koskinen}, D.~J. and {Kowalski}, M. and {Kowarik}, T. and {Krasberg}, M. and {Krings}, T. and {Kroll}, G. and {Kuehn}, K. and {Kuwabara}, T. and {Labare}, M. and {Lafebre}, S. and {Laihem}, K. and {Landsman}, H. and {Lauer}, R. and {Lehmann}, R. and {Lennarz}, D. and {L{\"u}nemann}, J. and {Madsen}, J. and {Majumdar}, P. and {Marotta}, A. and {Maruyama}, R. and {Mase}, K. and {Matis}, H.~S. and {Matusik}, M. and {Meagher}, K. and {Merck}, M. and {M{\'e}sz{\'a}ros}, P. and {Meures}, T. and {Middell}, E. and {Milke}, N. and {Miller}, J. and {Montaruli}, T. and {Morse}, R. and {Movit}, S.~M. and {Nahnhauer}, R. and {Nam}, J.~W. and {Naumann}, U. and {Nie{\ss}en}, P. and {Nygren}, D.~R. and {Odrowski}, S. and {Olivas}, A. and {Olivo}, M. and {O'Murchadha}, A. and {Ono}, M. and {Panknin}, S. and {Paul}, L. and {P{\'e}rez de los Heros}, C. and {Petrovic}, J. and {Piegsa}, A. and {Pieloth}, D. and {Porrata}, R. and {Posselt}, J. and {Price}, P.~B. and {Prikockis}, M. and {Przybylski}, G.~T. and {Rawlins}, K. and {Redl}, P. and {Resconi}, E. and {Rhode}, W. and {Ribordy}, M. and {Rizzo}, A. and {Rodrigues}, J.~P. and {Roth}, P. and {Rothmaier}, F. and {Rott}, C. and {Roucelle}, C. and {Ruhe}, T. and {Rutledge}, D. and {Ruzybayev}, B. and {Ryckbosch}, D. and {Sander}, H.-G. and {Santander}, M. and {Sarkar}, S. and {Schatto}, K. and {Schlenstedt}, S. and {Schmidt}, T. and {Schukraft}, A. and {Schultes}, A.},
        title = "{Measurement of the Anisotropy of Cosmic-ray Arrival Directions with IceCube}",
      journal = {Astrophys. J. Lett.},
         year = 2010,
        month = aug,
       volume = {718},
       number = {2},
        pages = {L194-L198},
          doi = {10.1088/2041-8205/718/2/L194},
archivePrefix = {arXiv},
       eprint = {1005.2960},
 primaryClass = {astro-ph.HE},
       adsurl = {https://ui.adsabs.harvard.edu/abs/2010ApJ...718L.194A}
}

@ARTICLE{2010ApJ...711..119A,
       author = {{Amenomori}, M. and {Bi}, X.~J. and {Chen}, D. and {Cui}, S.~W. and {Danzengluobu} and {Ding}, L.~K. and {Ding}, X.~H. and {Fan}, C. and {Feng}, C.~F. and {Feng}, Zhaoyang and {Feng}, Z.~Y. and {Gao}, X.~Y. and {Geng}, Q.~X. and {Gou}, Q.~B. and {Guo}, H.~W. and {He}, H.~H. and {He}, M. and {Hibino}, K. and {Hotta}, N. and {Hu}, Haibing and {Hu}, H.~B. and {Huang}, J. and {Huang}, Q. and {Jia}, H.~Y. and {Jiang}, L. and {Kajino}, F. and {Kasahara}, K. and {Katayose}, Y. and {Kato}, C. and {Kawata}, K. and {Labaciren} and {Le}, G.~M. and {Li}, A.~F. and {Li}, H.~C. and {Li}, J.~Y. and {Liu}, C. and {Lou}, Y.-Q. and {Lu}, H. and {Meng}, X.~R. and {Mizutani}, K. and {Mu}, J. and {Munakata}, K. and {Nagai}, A. and {Nanjo}, H. and {Nishizawa}, M. and {Ohnishi}, M. and {Ohta}, I. and {Ozawa}, S. and {Saito}, T. and {Saito}, T.~Y. and {Sakata}, M. and {Sako}, T.~K. and {Shibata}, M. and {Shiomi}, A. and {Shirai}, T. and {Sugimoto}, H. and {Takita}, M. and {Tan}, Y.~H. and {Tateyama}, N. and {Torii}, S. and {Tsuchiya}, H. and {Udo}, S. and {Wang}, B. and {Wang}, H. and {Wang}, Y. and {Wang}, Y.~G. and {Wu}, H.~R. and {Xue}, L. and {Yamamoto}, Y. and {Yan}, C.~T. and {Yang}, X.~C. and {Yasue}, S. and {Ye}, Z.~H. and {Yu}, G.~C. and {Yuan}, A.~F. and {Yuda}, T. and {Zhang}, H.~M. and {Zhang}, J.~L. and {Zhang}, N.~J. and {Zhang}, X.~Y. and {Zhang}, Y. and {Zhang}, Yi and {Zhang}, Ying and {Zhaxisangzhu} and {Zhou}, X.~X. and {Tibet AS{\ensuremath{\gamma}} Collaboration}},
        title = "{On Temporal Variations of the Multi-TeV Cosmic Ray Anisotropy Using the Tibet III Air Shower Array}",
      journal = {\apj},
         year = 2010,
        month = mar,
       volume = {711},
       number = {1},
        pages = {119-124},
          doi = {10.1088/0004-637X/711/1/119},
archivePrefix = {arXiv},
       eprint = {1001.2646},
 primaryClass = {astro-ph.HE},
       adsurl = {https://ui.adsabs.harvard.edu/abs/2010ApJ...711..119A}
}

@ARTICLE{2018ApJ...861...93B,
       author = {{Bartoli}, B. and {Bernardini}, P. and {Bi}, X.~J. and {Cao}, Z. and {Catalanotti}, S. and {Chen}, S.~Z. and {Chen}, T.~L. and {Cui}, S.~W. and {Dai}, B.~Z. and {D'Amone}, A. and {Danzengluobu} and {De Mitri}, I. and {D'Ettorre Piazzoli}, B. and {Di Girolamo}, T. and {Di Sciascio}, G. and {Feng}, C.~F. and {Feng}, Z.~Y. and {Gao}, W. and {Gou}, Q.~B. and {Guo}, Y.~Q. and {He}, H.~H. and {Hu}, Haibing and {Hu}, Hongbo and {Iacovacci}, M. and {Iuppa}, R. and {Jia}, H.~Y. and {Labaciren} and {Li}, H.~J. and {Liu}, C. and {Liu}, J. and {Liu}, M.~Y. and {Lu}, H. and {Ma}, L.~L. and {Ma}, X.~H. and {Mancarella}, G. and {Mari}, S.~M. and {Marsella}, G. and {Mastroianni}, S. and {Montini}, P. and {Ning}, C.~C. and {Perrone}, L. and {Pistilli}, P. and {Ruffolo}, D. and {Salvini}, P. and {Santonico}, R. and {Shen}, P.~R. and {Sheng}, X.~D. and {Shi}, F. and {Surdo}, A. and {Tan}, Y.~H. and {Vallania}, P. and {Vernetto}, S. and {Vigorito}, C. and {Wang}, H. and {Wu}, C.~Y. and {Wu}, H.~R. and {Xue}, L. and {Yang}, Q.~Y. and {Yang}, X.~C. and {Yao}, Z.~G. and {Yuan}, A.~F. and {Zha}, M. and {Zhang}, H.~M. and {Zhang}, L. and {Zhang}, X.~Y. and {Zhang}, Y. and {Zhao}, J. and {Zhaxiciren} and {Zhaxisangzhu} and {Zhou}, X.~X. and {Zhu}, F.~R. and {Zhu}, Q.~Q. and {ARGO-YBJ Collaboration}},
        title = "{Galactic Cosmic-Ray Anisotropy in the Northern Hemisphere from the ARGO-YBJ Experiment during 2008-2012}",
      journal = {\apj},
         year = 2018,
        month = jul,
       volume = {861},
       number = {2},
          eid = {93},
        pages = {93},
          doi = {10.3847/1538-4357/aac6cc},
archivePrefix = {arXiv},
       eprint = {1805.08980},
 primaryClass = {astro-ph.HE},
       adsurl = {https://ui.adsabs.harvard.edu/abs/2018ApJ...861...93B}
}

@ARTICLE{2011ApJ...740...16A,
       author = {{Abbasi}, R. and {Abdou}, Y. and {Abu-Zayyad}, T. and {Adams}, J. and {Aguilar}, J.~A. and {Ahlers}, M. and {Altmann}, D. and {Andeen}, K. and {Auffenberg}, J. and {Bai}, X. and {Baker}, M. and {Barwick}, S.~W. and {Bay}, R. and {Bazo Alba}, J.~L. and {Beattie}, K. and {Beatty}, J.~J. and {Bechet}, S. and {Becker}, J.~K. and {Becker}, K.-H. and {Benabderrahmane}, M.~L. and {BenZvi}, S. and {Berdermann}, J. and {Berghaus}, P. and {Berley}, D. and {Bernardini}, E. and {Bertrand}, D. and {Besson}, D.~Z. and {Bindig}, D. and {Bissok}, M. and {Blaufuss}, E. and {Blumenthal}, J. and {Boersma}, D.~J. and {Bohm}, C. and {Bose}, D. and {B{\"o}ser}, S. and {Botner}, O. and {Brown}, A.~M. and {Buitink}, S. and {Caballero-Mora}, K.~S. and {Carson}, M. and {Chirkin}, D. and {Christy}, B. and {Clem}, J. and {Clevermann}, F. and {Cohen}, S. and {Colnard}, C. and {Cowen}, D.~F. and {D'Agostino}, M.~V. and {Danninger}, M. and {Daughhetee}, J. and {Davis}, J.~C. and {De Clercq}, C. and {Demir{\"o}rs}, L. and {Denger}, T. and {Depaepe}, O. and {Descamps}, F. and {Desiati}, P. and {de Vries-Uiterweerd}, G. and {DeYoung}, T. and {D{\'\i}az-V{\'e}lez}, J.~C. and {Dierckxsens}, M. and {Dreyer}, J. and {Dumm}, J.~P. and {Ehrlich}, R. and {Eisch}, J. and {Ellsworth}, R.~W. and {Engdeg{\r{a}}rd}, O. and {Euler}, S. and {Evenson}, P.~A. and {Fadiran}, O. and {Fazely}, A.~R. and {Fedynitch}, A. and {Feintzeig}, J. and {Feusels}, T. and {Filimonov}, K. and {Finley}, C. and {Fischer-Wasels}, T. and {Foerster}, M.~M. and {Fox}, B.~D. and {Franckowiak}, A. and {Franke}, R. and {Gaisser}, T.~K. and {Gallagher}, J. and {Gerhardt}, L. and {Gladstone}, L. and {Gl{\"u}senkamp}, T. and {Goldschmidt}, A. and {Goodman}, J.~A. and {Gora}, D. and {Grant}, D. and {Griesel}, T. and {Gro{\ss}}, A. and {Grullon}, S. and {Gurtner}, M. and {Ha}, C. and {Hajismail}, A. and {Hallgren}, A. and {Halzen}, F. and {Han}, K. and {Hanson}, K. and {Heinen}, D. and {Helbing}, K. and {Herquet}, P. and {Hickford}, S. and {Hill}, G.~C. and {Hoffman}, K.~D. and {Homeier}, A. and {Hoshina}, K. and {Hubert}, D. and {Huelsnitz}, W. and {H{\"u}l{\ss}}, J.-P. and {Hulth}, P.~O. and {Hultqvist}, K. and {Hussain}, S. and {Ishihara}, A. and {Jacobsen}, J. and {Japaridze}, G.~S. and {Johansson}, H. and {Joseph}, J.~M. and {Kampert}, K.-H. and {Kappes}, A. and {Karg}, T. and {Karle}, A. and {Kenny}, P. and {Kiryluk}, J. and {Kislat}, F. and {Klein}, S.~R. and {K{\"o}hne}, J.-H. and {Kohnen}, G. and {Kolanoski}, H. and {K{\"o}pke}, L. and {Kopper}, S. and {Koskinen}, D.~J. and {Kowalski}, M. and {Kowarik}, T. and {Krasberg}, M. and {Krings}, T. and {Kroll}, G. and {Kurahashi}, N. and {Kuwabara}, T. and {Labare}, M. and {Lafebre}, S. and {Laihem}, K. and {Landsman}, H. and {Larson}, M.~J. and {Lauer}, R. and {L{\"u}nemann}, J. and {Madajczyk}, B. and {Madsen}, J. and {Majumdar}, P. and {Marotta}, A. and {Maruyama}, R. and {Mase}, K. and {Matis}, H.~S. and {Meagher}, K. and {Merck}, M. and {M{\'e}sz{\'a}ros}, P. and {Meures}, T. and {Middell}, E. and {Milke}, N. and {Miller}, J. and {Montaruli}, T. and {Morse}, R. and {Movit}, S.~M. and {Nahnhauer}, R. and {Nam}, J.~W. and {Naumann}, U. and {Nie{\ss}en}, P. and {Nygren}, D.~R. and {Odrowski}, S. and {Olivas}, A. and {Olivo}, M. and {O'Murchadha}, A. and {Ono}, M. and {Panknin}, S. and {Paul}, L. and {P{\'e}rez de los Heros}, C. and {Petrovic}, J. and {Piegsa}, A. and {Pieloth}, D. and {Porrata}, R. and {Posselt}, J. and {Price}, C.~C. and {Price}, P.~B. and {Przybylski}, G.~T. and {Rawlins}, K. and {Redl}, P. and {Resconi}, E. and {Rhode}, W. and {Ribordy}, M. and {Rizzo}, A. and {Rodrigues}, J.~P. and {Roth}, P. and {Rothmaier}, F. and {Rott}, C. and {Ruhe}, T. and {Rutledge}, D. and {Ruzybayev}, B. and {Ryckbosch}, D. and {Sander}, H.-G.},
        title = "{Observation of Anisotropy in the Arrival Directions of Galactic Cosmic Rays at Multiple Angular Scales with IceCube}",
      journal = {\apj},
         year = 2011,
        month = oct,
       volume = {740},
       number = {1},
          eid = {16},
        pages = {16},
          doi = {10.1088/0004-637X/740/1/16},
archivePrefix = {arXiv},
       eprint = {1105.2326},
 primaryClass = {astro-ph.HE},
       adsurl = {https://ui.adsabs.harvard.edu/abs/2011ApJ...740...16A}
}

@ARTICLE{2012ApJ...746...33A,
       author = {{Abbasi}, R. and {Abdou}, Y. and {Abu-Zayyad}, T. and {Ackermann}, M. and {Adams}, J. and {Aguilar}, J.~A. and {Ahlers}, M. and {Allen}, M.~M. and {Altmann}, D. and {Andeen}, K. and {Auffenberg}, J. and {Bai}, X. and {Baker}, M. and {Barwick}, S.~W. and {Bay}, R. and {Bazo Alba}, J.~L. and {Beattie}, K. and {Beatty}, J.~J. and {Bechet}, S. and {Becker}, J.~K. and {Becker}, K.-H. and {Benabderrahmane}, M.~L. and {BenZvi}, S. and {Berdermann}, J. and {Berghaus}, P. and {Berley}, D. and {Bernardini}, E. and {Bertrand}, D. and {Besson}, D.~Z. and {Bindig}, D. and {Bissok}, M. and {Blaufuss}, E. and {Blumenthal}, J. and {Boersma}, D.~J. and {Bohm}, C. and {Bose}, D. and {B{\"o}ser}, S. and {Botner}, O. and {Brown}, A.~M. and {Buitink}, S. and {Caballero-Mora}, K.~S. and {Carson}, M. and {Chirkin}, D. and {Christy}, B. and {Clevermann}, F. and {Cohen}, S. and {Colnard}, C. and {Cowen}, D.~F. and {Cruz Silva}, A.~H. and {D'Agostino}, M.~V. and {Danninger}, M. and {Daughhetee}, J. and {Davis}, J.~C. and {De Clercq}, C. and {Degner}, T. and {Demir{\"o}rs}, L. and {Descamps}, F. and {Desiati}, P. and {de Vries-Uiterweerd}, G. and {DeYoung}, T. and {D{\'\i}az-V{\'e}lez}, J.~C. and {Dierckxsens}, M. and {Dreyer}, J. and {Dumm}, J.~P. and {Dunkman}, M. and {Eisch}, J. and {Ellsworth}, R.~W. and {Engdeg{\r{a}}rd}, O. and {Euler}, S. and {Evenson}, P.~A. and {Fadiran}, O. and {Fazely}, A.~R. and {Fedynitch}, A. and {Feintzeig}, J. and {Feusels}, T. and {Filimonov}, K. and {Finley}, C. and {Fischer-Wasels}, T. and {Fox}, B.~D. and {Franckowiak}, A. and {Franke}, R. and {Gaisser}, T.~K. and {Gallagher}, J. and {Gerhardt}, L. and {Gladstone}, L. and {Gl{\"u}senkamp}, T. and {Goldschmidt}, A. and {Goodman}, J.~A. and {G{\'o}ra}, D. and {Grant}, D. and {Griesel}, T. and {Gro{\ss}}, A. and {Grullon}, S. and {Gurtner}, M. and {Ha}, C. and {Haj Ismail}, A. and {Hallgren}, A. and {Halzen}, F. and {Han}, K. and {Hanson}, K. and {Heinen}, D. and {Helbing}, K. and {Hellauer}, R. and {Hickford}, S. and {Hill}, G.~C. and {Hoffman}, K.~D. and {Hoffmann}, B. and {Homeier}, A. and {Hoshina}, K. and {Huelsnitz}, W. and {H{\"u}l{\ss}}, J.-P. and {Hulth}, P.~O. and {Hultqvist}, K. and {Hussain}, S. and {Ishihara}, A. and {Jacobi}, E. and {Jacobsen}, J. and {Japaridze}, G.~S. and {Johansson}, H. and {Kampert}, K.-H. and {Kappes}, A. and {Karg}, T. and {Karle}, A. and {Kenny}, P. and {Kiryluk}, J. and {Kislat}, F. and {Klein}, S.~R. and {K{\"o}hne}, J.-H. and {Kohnen}, G. and {Kolanoski}, H. and {K{\"o}pke}, L. and {Kopper}, S. and {Koskinen}, D.~J. and {Kowalski}, M. and {Kowarik}, T. and {Krasberg}, M. and {Kroll}, G. and {Kurahashi}, N. and {Kuwabara}, T. and {Labare}, M. and {Laihem}, K. and {Landsman}, H. and {Larson}, M.~J. and {Lauer}, R. and {L{\"u}nemann}, J. and {Madsen}, J. and {Marotta}, A. and {Maruyama}, R. and {Mase}, K. and {Matis}, H.~S. and {Meagher}, K. and {Merck}, M. and {M{\'e}sz{\'a}ros}, P. and {Meures}, T. and {Miarecki}, S. and {Middell}, E. and {Milke}, N. and {Miller}, J. and {Montaruli}, T. and {Morse}, R. and {Movit}, S.~M. and {Nahnhauer}, R. and {Nam}, J.~W. and {Naumann}, U. and {Nygren}, D.~R. and {Odrowski}, S. and {Olivas}, A. and {Olivo}, M. and {O'Murchadha}, A. and {Panknin}, S. and {Paul}, L. and {P{\'e}rez de los Heros}, C. and {Petrovic}, J. and {Piegsa}, A. and {Pieloth}, D. and {Porrata}, R. and {Posselt}, J. and {Price}, C.~C. and {Price}, P.~B. and {Przybylski}, G.~T. and {Rawlins}, K. and {Redl}, P. and {Resconi}, E. and {Rhode}, W. and {Ribordy}, M. and {Richman}, M. and {Rodrigues}, J.~P. and {Rothmaier}, F. and {Rott}, C. and {Ruhe}, T. and {Rutledge}, D. and {Ruzybayev}, B. and {Ryckbosch}, D. and {Sander}, H.-G. and {Santander}, M. and {Sarkar}, S. and {Schatto}, K. and {Schmidt}, T. and {Sch{\"o}nwald}, A. and {Schukraft}, A.},
        title = "{Observation of Anisotropy in the Galactic Cosmic-Ray Arrival Directions at 400 TeV with IceCube}",
      journal = {\apj},
         year = 2012,
        month = feb,
       volume = {746},
       number = {1},
          eid = {33},
        pages = {33},
          doi = {10.1088/0004-637X/746/1/33},
archivePrefix = {arXiv},
       eprint = {1109.1017},
 primaryClass = {hep-ex},
       adsurl = {https://ui.adsabs.harvard.edu/abs/2012ApJ...746...33A}
}

@ARTICLE{2012APh....39..144L,
       author = {{Li}, Taoli and {Liu}, Maoyuan and {Cui}, Shuwang and {Hou}, Zhengtao},
        title = "{Evaluation of a wide-sky survey method for EAS experiments}",
      journal = {Astroparticle Physics},
         year = 2012,
        month = dec,
       volume = {39},
        pages = {144-148},
          doi = {10.1016/j.astropartphys.2012.11.001},
       adsurl = {https://ui.adsabs.harvard.edu/abs/2012APh....39..144L}
}

@ARTICLE{2013ApJ...765...55A,
       author = {{Aartsen}, M.~G. and {Abbasi}, R. and {Abdou}, Y. and {Ackermann}, M. and {Adams}, J. and {Aguilar}, J.~A. and {Ahlers}, M. and {Altmann}, D. and {Andeen}, K. and {Auffenberg}, J. and {Bai}, X. and {Baker}, M. and {Barwick}, S.~W. and {Baum}, V. and {Bay}, R. and {Beattie}, K. and {Beatty}, J.~J. and {Bechet}, S. and {Becker Tjus}, J. and {Becker}, K.-H. and {Bell}, M. and {Benabderrahmane}, M.~L. and {BenZvi}, S. and {Berdermann}, J. and {Berghaus}, P. and {Berley}, D. and {Bernardini}, E. and {Bertrand}, D. and {Besson}, D.~Z. and {Bindig}, D. and {Bissok}, M. and {Blaufuss}, E. and {Blumenthal}, J. and {Boersma}, D.~J. and {Bohaichuk}, S. and {Bohm}, C. and {Bose}, D. and {B{\"o}ser}, S. and {Botner}, O. and {Brayeur}, L. and {Brown}, A.~M. and {Bruijn}, R. and {Brunner}, J. and {Carson}, M. and {Casey}, J. and {Casier}, M. and {Chirkin}, D. and {Christy}, B. and {Clark}, K. and {Clevermann}, F. and {Cohen}, S. and {Cowen}, D.~F. and {Cruz Silva}, A.~H. and {Danninger}, M. and {Daughhetee}, J. and {Davis}, J.~C. and {De Clercq}, C. and {De Ridder}, S. and {Descamps}, F. and {Desiati}, P. and {de Vries-Uiterweerd}, G. and {DeYoung}, T. and {D{\'\i}az-V{\'e}lez}, J.~C. and {Dreyer}, J. and {Dumm}, J.~P. and {Dunkman}, M. and {Eagan}, R. and {Eisch}, J. and {Ellsworth}, R.~W. and {Engdeg{\r{a}}rd}, O. and {Euler}, S. and {Evenson}, P.~A. and {Fadiran}, O. and {Fazely}, A.~R. and {Fedynitch}, A. and {Feintzeig}, J. and {Feusels}, T. and {Filimonov}, K. and {Finley}, C. and {Fischer-Wasels}, T. and {Flis}, S. and {Franckowiak}, A. and {Franke}, R. and {Frantzen}, K. and {Fuchs}, T. and {Gaisser}, T.~K. and {Gallagher}, J. and {Gerhardt}, L. and {Gladstone}, L. and {Gl{\"u}senkamp}, T. and {Goldschmidt}, A. and {Golup}, G. and {Goodman}, J.~A. and {G{\'o}ra}, D. and {Grant}, D. and {Gross}, A. and {Grullon}, S. and {Gurtner}, M. and {Ha}, C. and {Haj Ismail}, A. and {Hallgren}, A. and {Halzen}, F. and {Hanson}, K. and {Heereman}, D. and {Heimann}, P. and {Heinen}, D. and {Helbing}, K. and {Hellauer}, R. and {Hickford}, S. and {Hill}, G.~C. and {Hoffman}, K.~D. and {Hoffmann}, R. and {Homeier}, A. and {Hoshina}, K. and {Huelsnitz}, W. and {Hulth}, P.~O. and {Hultqvist}, K. and {Hussain}, S. and {Ishihara}, A. and {Jacobi}, E. and {Jacobsen}, J. and {Japaridze}, G.~S. and {Jlelati}, O. and {Kappes}, A. and {Karg}, T. and {Karle}, A. and {Kiryluk}, J. and {Kislat}, F. and {Kl{\"a}s}, J. and {Klein}, S.~R. and {K{\"o}hne}, J.-H. and {Kohnen}, G. and {Kolanoski}, H. and {K{\"o}pke}, L. and {Kopper}, C. and {Kopper}, S. and {Koskinen}, D.~J. and {Kowalski}, M. and {Krasberg}, M. and {Kroll}, G. and {Kunnen}, J. and {Kurahashi}, N. and {Kuwabara}, T. and {Labare}, M. and {Landsman}, H. and {Larson}, M.~J. and {Lauer}, R. and {Lesiak-Bzdak}, M. and {L{\"u}nemann}, J. and {Madsen}, J. and {Maruyama}, R. and {Mase}, K. and {Matis}, H.~S. and {McNally}, F. and {Meagher}, K. and {Merck}, M. and {M{\'e}sz{\'a}ros}, P. and {Meures}, T. and {Miarecki}, S. and {Middell}, E. and {Milke}, N. and {Miller}, J. and {Mohrmann}, L. and {Montaruli}, T. and {Morse}, R. and {Nahnhauer}, R. and {Naumann}, U. and {Nowicki}, S.~C. and {Nygren}, D.~R. and {Obertacke}, A. and {Odrowski}, S. and {Olivas}, A. and {Olivo}, M. and {O'Murchadha}, A. and {Panknin}, S. and {Paul}, L. and {Pepper}, J.~A. and {P{\'e}rez de los Heros}, C. and {Pieloth}, D. and {Pirk}, N. and {Posselt}, J. and {Price}, P.~B. and {Przybylski}, G.~T. and {R{\"a}del}, L. and {Rawlins}, K. and {Redl}, P. and {Resconi}, E. and {Rhode}, W. and {Ribordy}, M. and {Richman}, M. and {Riedel}, B. and {Rodrigues}, J.~P. and {Rothmaier}, F. and {Rott}, C. and {Ruhe}, T. and {Ruzybayev}, B. and {Ryckbosch}, D. and {Saba}, S.~M. and {Salameh}, T. and {Sander}, H.-G.},
        title = "{Observation of Cosmic-Ray Anisotropy with the IceTop Air Shower Array}",
      journal = {\apj},
         year = 2013,
        month = mar,
       volume = {765},
       number = {1},
          eid = {55},
        pages = {55},
          doi = {10.1088/0004-637X/765/1/55},
archivePrefix = {arXiv},
       eprint = {1210.5278},
 primaryClass = {astro-ph.HE},
       adsurl = {https://ui.adsabs.harvard.edu/abs/2013ApJ...765...55A}
}

@ARTICLE{2013PhRvD..88h2001B,
       author = {{Bartoli}, B. and {Bernardini}, P. and {Bi}, X.~J. and {Bolognino}, I. and {Branchini}, P. and {Budano}, A. and {Calabrese Melcarne}, A.~K. and {Camarri}, P. and {Cao}, Z. and {Cardarelli}, R. and {Catalanotti}, S. and {Chen}, S.~Z. and {Chen}, T.~L. and {Creti}, P. and {Cui}, S.~W. and {Dai}, B.~Z. and {D'Amone}, A. and {Danzengluobu} and {De Mitri}, I. and {D'Ettorre Piazzoli}, B. and {Di Girolamo}, T. and {Di Sciascio}, G. and {Feng}, C.~F. and {Feng}, Zhaoyang and {Feng}, Zhenyong and {Gou}, Q.~B. and {Guo}, Y.~Q. and {He}, H.~H. and {Hu}, Haibing and {Hu}, Hongbo and {Iacovacci}, M. and {Iuppa}, R. and {Jia}, H.~Y. and {Labaciren} and {Li}, H.~J. and {Liguori}, G. and {Liu}, C. and {Liu}, J. and {Liu}, M.~Y. and {Lu}, H. and {Ma}, X.~H. and {Mancarella}, G. and {Mari}, S.~M. and {Marsella}, G. and {Martello}, D. and {Mastroianni}, S. and {Montini}, P. and {Ning}, C.~C. and {Panareo}, M. and {Panico}, B. and {Perrone}, L. and {Pistilli}, P. and {Ruggieri}, F. and {Salvini}, P. and {Santonico}, R. and {Sbano}, S.~N. and {Shen}, P.~R. and {Sheng}, X.~D. and {Shi}, F. and {Surdo}, A. and {Tan}, Y.~H. and {Vallania}, P. and {Vernetto}, S. and {Vigorito}, C. and {Wang}, H. and {Wu}, C.~Y. and {Wu}, H.~R. and {Xue}, L. and {Yan}, Y.~X. and {Yang}, Q.~Y. and {Yang}, X.~C. and {Yao}, Z.~G. and {Yuan}, A.~F. and {Zha}, M. and {Zhang}, H.~M. and {Zhang}, L. and {Zhang}, X.~Y. and {Zhang}, Y. and {Zhaxiciren} and {Zhaxisangzhu} and {Zhou}, X.~X. and {Zhu}, F.~R. and {Zhu}, Q.~Q. and {Zizzi}, G.},
        title = "{Medium scale anisotropy in the TeV cosmic ray flux observed by ARGO-YBJ}",
      journal = {\prd},
         year = 2013,
        month = oct,
       volume = {88},
       number = {8},
          eid = {082001},
        pages = {082001},
          doi = {10.1103/PhysRevD.88.082001},
archivePrefix = {arXiv},
       eprint = {1309.6182},
 primaryClass = {astro-ph.HE},
       adsurl = {https://ui.adsabs.harvard.edu/abs/2013PhRvD..88h2001B}
}

@ARTICLE{2015ApJ...809...90B,
       author = {{Bartoli}, B. and {Bernardini}, P. and {Bi}, X.~J. and {Cao}, Z. and {Catalanotti}, S. and {Chen}, S.~Z. and {Chen}, T.~L. and {Cui}, S.~W. and {Dai}, B.~Z. and {D'Amone}, A. and {Danzengluobu} and {De Mitri}, I. and {D'Ettorre Piazzoli}, B. and {Di Girolamo}, T. and {Di Sciascio}, G. and {Feng}, C.~F. and {Feng}, Zhaoyang and {Feng}, Zhenyong and {Gao}, W. and {Gou}, Q.~B. and {Guo}, Y.~Q. and {He}, H.~H. and {Hu}, Haibing and {Hu}, Hongbo and {Iacovacci}, M. and {Iuppa}, R. and {Jia}, H.~Y. and {Labaciren} and {Li}, H.~J. and {Liu}, C. and {Liu}, J. and {Liu}, M.~Y. and {Lu}, H. and {Ma}, L.~L. and {Ma}, X.~H. and {Mancarella}, G. and {Mari}, S.~M. and {Marsella}, G. and {Mastroianni}, S. and {Montini}, P. and {Ning}, C.~C. and {Perrone}, L. and {Pistilli}, P. and {Salvini}, P. and {Santonico}, R. and {Shen}, P.~R. and {Sheng}, X.~D. and {Shi}, F. and {Surdo}, A. and {Tan}, Y.~H. and {Vallania}, P. and {Vernetto}, S. and {Vigorito}, C. and {Wang}, H. and {Wu}, C.~Y. and {Wu}, H.~R. and {Xue}, L. and {Yang}, Q.~Y. and {Yang}, X.~C. and {Yao}, Z.~G. and {Yuan}, A.~F. and {Zha}, M. and {Zhang}, H.~M. and {Zhang}, L. and {Zhang}, X.~Y. and {Zhang}, Y. and {Zhao}, J. and {Zhaxiciren} and {Zhaxisangzhu} and {Zhou}, X.~X. and {Zhu}, F.~R. and {Zhu}, Q.~Q. and {ARGO-YBJ Collaboration}},
        title = "{ARGO-YBJ Observation of the Large-scale Cosmic Ray Anisotropy During the Solar Minimum between Cycles 23 and 24}",
      journal = {\apj},
         year = 2015,
        month = aug,
       volume = {809},
       number = {1},
          eid = {90},
        pages = {90},
          doi = {10.1088/0004-637X/809/1/90},
archivePrefix = {arXiv},
       eprint = {2112.14891},
 primaryClass = {astro-ph.HE},
       adsurl = {https://ui.adsabs.harvard.edu/abs/2015ApJ...809...90B}
}

@ARTICLE{2014ApJ...796..108A,
       author = {{Abeysekara}, A.~U. and {Alfaro}, R. and {Alvarez}, C. and {{\'A}lvarez}, J.~D. and {Arceo}, R. and {Arteaga-Vel{\'a}zquez}, J.~C. and {Ayala Solares}, H.~A. and {Barber}, A.~S. and {Baughman}, B.~M. and {Bautista-Elivar}, N. and {Belmont}, E. and {BenZvi}, S.~Y. and {Berley}, D. and {Bonilla Rosales}, M. and {Braun}, J. and {Caballero-Mora}, K.~S. and {Carrami{\~n}ana}, A. and {Castillo}, M. and {Cotti}, U. and {Cotzomi}, J. and {de la Fuente}, E. and {De Le{\'o}n}, C. and {DeYoung}, T. and {Diaz Hernandez}, R. and {D{\'\i}az-V{\'e}lez}, J.~C. and {Dingus}, B.~L. and {DuVernois}, M.~A. and {Ellsworth}, R.~W. and {Fiorino}, D.~W. and {Fraija}, N. and {Galindo}, A. and {Garfias}, F. and {Gonz{\'a}lez}, M.~M. and {Goodman}, J.~A. and {Gussert}, M. and {Hampel-Arias}, Z. and {Harding}, J.~P. and {H{\"u}ntemeyer}, P. and {Hui}, C.~M. and {Imran}, A. and {Iriarte}, A. and {Karn}, P. and {Kieda}, D. and {Kunde}, G.~J. and {Lara}, A. and {Lauer}, R.~J. and {Lee}, W.~H. and {Lennarz}, D. and {Le{\'o}n Vargas}, H. and {Linnemann}, J.~T. and {Longo}, M. and {Luna-Garc{\'\i}a}, R. and {Malone}, K. and {Marinelli}, A. and {Marinelli}, S.~S. and {Martinez}, H. and {Martinez}, O. and {Mart{\'\i}nez-Castro}, J. and {Matthews}, J.~A.~J. and {McEnery}, J. and {Mendoza Torres}, E. and {Miranda-Romagnoli}, P. and {Moreno}, E. and {Mostaf{\'a}}, M. and {Nellen}, L. and {Newbold}, M. and {Noriega-Papaqui}, R. and {Oceguera-Becerra}, T. and {Patricelli}, B. and {Pelayo}, R. and {P{\'e}rez-P{\'e}rez}, E.~G. and {Pretz}, J. and {Rivi{\`e}re}, C. and {Rosa-Gonz{\'a}lez}, D. and {Ruiz-Velasco}, E. and {Ryan}, J. and {Salazar}, H. and {Salesa Greus}, F. and {Sandoval}, A. and {Schneider}, M. and {Sinnis}, G. and {Smith}, A.~J. and {Sparks Woodle}, K. and {Springer}, R.~W. and {Taboada}, I. and {Toale}, P.~A. and {Tollefson}, K. and {Torres}, I. and {Ukwatta}, T.~N. and {Villase{\~n}or}, L. and {Weisgarber}, T. and {Westerhoff}, S. and {Wisher}, I.~G. and {Wood}, J. and {Yodh}, G.~B. and {Younk}, P.~W. and {Zaborov}, D. and {Zepeda}, A. and {Zhou}, H. and {HAWC Collaboration}},
        title = "{Observation of Small-scale Anisotropy in the Arrival Direction Distribution of TeV Cosmic Rays with HAWC}",
      journal = {\apj},
         year = 2014,
        month = dec,
       volume = {796},
       number = {2},
          eid = {108},
        pages = {108},
          doi = {10.1088/0004-637X/796/2/108},
archivePrefix = {arXiv},
       eprint = {1408.4805},
 primaryClass = {astro-ph.HE},
       adsurl = {https://ui.adsabs.harvard.edu/abs/2014ApJ...796..108A}
}

@ARTICLE{2016ApJ...826..220A,
       author = {{Aartsen}, M.~G. and {Abraham}, K. and {Ackermann}, M. and {Adams}, J. and {Aguilar}, J.~A. and {Ahlers}, M. and {Ahrens}, M. and {Altmann}, D. and {Anderson}, T. and {Ansseau}, I. and {Anton}, G. and {Archinger}, M. and {Arguelles}, C. and {Arlen}, T.~C. and {Auffenberg}, J. and {Bai}, X. and {Barwick}, S.~W. and {Baum}, V. and {Bay}, R. and {Beatty}, J.~J. and {Becker Tjus}, J. and {Becker}, K.-H. and {Beiser}, E. and {BenZvi}, S. and {Berghaus}, P. and {Berley}, D. and {Bernardini}, E. and {Bernhard}, A. and {Besson}, D.~Z. and {Binder}, G. and {Bindig}, D. and {Bissok}, M. and {Blaufuss}, E. and {Blumenthal}, J. and {Boersma}, D.~J. and {Bohm}, C. and {B{\"o}rner}, M. and {Bos}, F. and {Bose}, D. and {B{\"o}ser}, S. and {Botner}, O. and {Braun}, J. and {Brayeur}, L. and {Bretz}, H.-P. and {Buzinsky}, N. and {Casey}, J. and {Casier}, M. and {Cheung}, E. and {Chirkin}, D. and {Christov}, A. and {Clark}, K. and {Classen}, L. and {Coenders}, S. and {Collin}, G.~H. and {Conrad}, J.~M. and {Cowen}, D.~F. and {Cruz Silva}, A.~H. and {Daughhetee}, J. and {Davis}, J.~C. and {Day}, M. and {de Andr{\'e}}, J.~P.~A.~M. and {De Clercq}, C. and {del Pino Rosendo}, E. and {Dembinski}, H. and {De Ridder}, S. and {Desiati}, P. and {de Vries}, K.~D. and {de Wasseige}, G. and {de With}, M. and {DeYoung}, T. and {D{\'\i}az-V{\'e}lez}, J.~C. and {di Lorenzo}, V. and {Dujmovic}, H. and {Dumm}, J.~P. and {Dunkman}, M. and {Eberhardt}, B. and {Ehrhardt}, T. and {Eichmann}, B. and {Euler}, S. and {Evenson}, P.~A. and {Fahey}, S. and {Fazely}, A.~R. and {Feintzeig}, J. and {Felde}, J. and {Filimonov}, K. and {Finley}, C. and {Flis}, S. and {F{\"o}sig}, C.-C. and {Fuchs}, T. and {Gaisser}, T.~K. and {Gaior}, R. and {Gallagher}, J. and {Gerhardt}, L. and {Ghorbani}, K. and {Gier}, D. and {Gladstone}, L. and {Glagla}, M. and {Gl{\"u}senkamp}, T. and {Goldschmidt}, A. and {Golup}, G. and {Gonzalez}, J.~G. and {G{\'o}ra}, D. and {Grant}, D. and {Griffith}, Z. and {Ha}, C. and {Haack}, C. and {Haj Ismail}, A. and {Hallgren}, A. and {Halzen}, F. and {Hansen}, E. and {Hansmann}, B. and {Hansmann}, T. and {Hanson}, K. and {Hebecker}, D. and {Heereman}, D. and {Helbing}, K. and {Hellauer}, R. and {Hickford}, S. and {Hignight}, J. and {Hill}, G.~C. and {Hoffman}, K.~D. and {Hoffmann}, R. and {Holzapfel}, K. and {Homeier}, A. and {Hoshina}, K. and {Huang}, F. and {Huber}, M. and {Huelsnitz}, W. and {Hulth}, P.~O. and {Hultqvist}, K. and {In}, S. and {Ishihara}, A. and {Jacobi}, E. and {Japaridze}, G.~S. and {Jeong}, M. and {Jero}, K. and {Jones}, B.~J.~P. and {Jurkovic}, M. and {Kappes}, A. and {Karg}, T. and {Karle}, A. and {Katz}, U. and {Kauer}, M. and {Keivani}, A. and {Kelley}, J.~L. and {Kemp}, J. and {Kheirandish}, A. and {Kim}, M. and {Kintscher}, T. and {Kiryluk}, J. and {Klein}, S.~R. and {Kohnen}, G. and {Koirala}, R. and {Kolanoski}, H. and {Konietz}, R. and {K{\"o}pke}, L. and {Kopper}, C. and {Kopper}, S. and {Koskinen}, D.~J. and {Kowalski}, M. and {Krings}, K. and {Kroll}, G. and {Kroll}, M. and {Kr{\"u}ckl}, G. and {Kunnen}, J. and {Kunwar}, S. and {Kurahashi}, N. and {Kuwabara}, T. and {Labare}, M. and {Lanfranchi}, J.~L. and {Larson}, M.~J. and {Lennarz}, D. and {Lesiak-Bzdak}, M. and {Leuermann}, M. and {Leuner}, J. and {Lu}, L. and {L{\"u}nemann}, J. and {Madsen}, J. and {Maggi}, G. and {Mahn}, K.~B.~M. and {Mandelartz}, M. and {Maruyama}, R. and {Mase}, K. and {Matis}, H.~S. and {Maunu}, R. and {McNally}, F. and {Meagher}, K. and {Medici}, M. and {Meier}, M. and {Meli}, A. and {Menne}, T. and {Merino}, G. and {Meures}, T. and {Miarecki}, S. and {Middell}, E. and {Mohrmann}, L. and {Montaruli}, T. and {Morse}, R. and {Nahnhauer}, R. and {Naumann}, U.},
        title = "{Anisotropy in Cosmic-Ray Arrival Directions in the Southern Hemisphere Based on Six Years of Data from the IceCube Detector}",
      journal = {\apj},
         year = 2016,
        month = aug,
       volume = {826},
       number = {2},
          eid = {220},
        pages = {220},
          doi = {10.3847/0004-637X/826/2/220},
archivePrefix = {arXiv},
       eprint = {1603.01227},
 primaryClass = {astro-ph.HE},
       adsurl = {https://ui.adsabs.harvard.edu/abs/2016ApJ...826..220A}
}

@ARTICLE{1935PhRv...47..817C,
       author = {{Compton}, Arthur H. and {Getting}, Ivan A.},
        title = "{An Apparent Effect of Galactic Rotation on the Intensity of Cosmic Rays}",
      journal = {Physical Review},
         year = 1935,
        month = jun,
       volume = {47},
       number = {11},
        pages = {817-821},
          doi = {10.1103/PhysRev.47.817},
       adsurl = {https://ui.adsabs.harvard.edu/abs/1935PhRv...47..817C}
}

@ARTICLE{2016PhRvL.117o1103A,
       author = {{Ahlers}, Markus},
        title = "{Deciphering the Dipole Anisotropy of Galactic Cosmic Rays}",
      journal = {\prl},
         year = 2016,
        month = oct,
       volume = {117},
       number = {15},
          eid = {151103},
        pages = {151103},
          doi = {10.1103/PhysRevLett.117.151103},
archivePrefix = {arXiv},
       eprint = {1605.06446},
 primaryClass = {astro-ph.HE},
       adsurl = {https://ui.adsabs.harvard.edu/abs/2016PhRvL.117o1103A}
}

@ARTICLE{1975PhRvL..34.1530L,
       author = {{Linsley}, John},
        title = "{Fluctuation effects on directional data}",
      journal = {\prl},
         year = 1975,
        month = jun,
       volume = {34},
       number = {24},
        pages = {1530-1533},
          doi = {10.1103/PhysRevLett.34.1530},
       adsurl = {https://ui.adsabs.harvard.edu/abs/1975PhRvL..34.1530L}
}

\end{document}